\documentclass[notoc]{JHEP}

\def\p{\partial}

\usepackage{epsfig}
\usepackage{amsmath}
\usepackage{amsfonts}
\usepackage{amssymb}

\title{Calculating the infra-red limit of quantum field theory 
using analyticity of correlation functions}
        
\author{Lars Kj\ae rgaard and Paul Mansfield \\ 
Department of Mathematical Sciences \\
 University of Durham \\  South Road, Durham DH1 3LE, U.K.\\
\email{lars.kjaergaard@durham.ac.uk} \\
\email{p.r.w.mansfield@durham.ac.uk}}

\abstract{
We describe a general method for calculating the infra-red limit of
physical quantities in unitary quantum field theories. Using  
analyticity of Green functions in a complex scale parameter, the
infra-red limit is expressed as a contour integral  entirely in the
ultra-violet region. 
The infra-red
limit is shown to be the limit of the Borel transform of the
physical quantity. 
The method is illustrated by
calculating the central charge of the perturbed unitary minimal models and the
critical exponents of $\varphi^4$ theory in three dimensions. 
We obtain approximate values for the
central charge which are 
very close to the exact values using only a one loop perturbative 
calculation. For $\varphi^4$ theory we obtain estimates  which are within the
errors of other more elaborate approaches.}

\preprint{DTP/99/82 \\ hep-th/9912048}

\begin{document}
\section{Introduction and summary}
It is important to get a better understanding of the infra-red limit of
quantum field theory. Conventionally this is studied by extrapolating 
perturbation theory from the ultra-violet using the renormalisation group.
The purpose of this paper is to show how to augment this approach with
information derived from the analyticity properties of Green functions.
These properties are well-known. 
For example,
the K\"{a}llen-Lehmann spectral representation of the two-point 
function of a scalar field shows that it has an analytic continuation to the
complex momentum plane. We will use this to express the infra-red limit as 
an integral
in the ultra-violet region where perturbation theory is applicable. 
For illustration assume that the physical quantity $F$ depends on 
some distance scale $s$, and that $F(s)$ is analytic in the
complex plane with the negative axis cut away, then the contour integral
\begin{equation}
  \label{firsteq}
  \frac{1}{2\pi i}\int_{C}\frac{ds}{s}e^{\rho/s}F(s)=0
\nonumber
\end{equation}
vanishes, where the contour $C$ is given in figure 1. 
This means that we can write the infra-red limit $F_{IR}=\lim_{|s|\rightarrow
  \infty}F(s)$ as an integral
\begin{equation}
  \label{secondeq}
  F_{IR}= \frac{1}{2\pi i}\left( \int_{C_0}+\int_{C_1}\right)\frac{ds}{s}e^{\rho/s}F(s)
\nonumber
\end{equation}
over the infinitesimal circle $C_0$ and the cut $C_1$. 

\begin{figure}
\epsfxsize=7cm
\epsfysize=7cm
\begin{center}
\epsffile{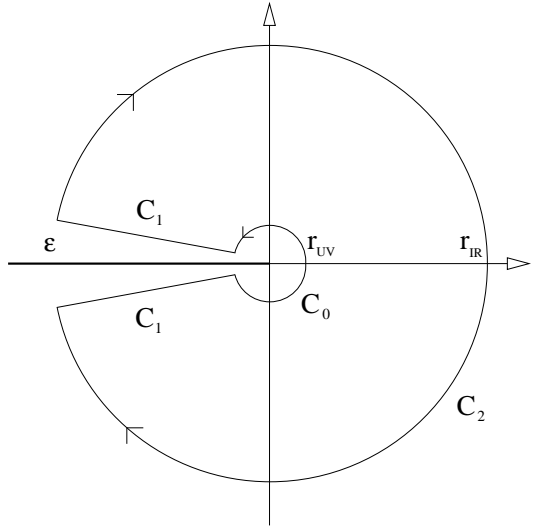}
\end{center}
\caption{Integration contour C in the cut complex plane.}
\label{fig:contour}
\end{figure}

We show below that the integral over the cut $C_1$ vanishes for large $\rho$,
so that the infra-red limit $F_{IR}$ then is given as an integral in the UV
region over $C_0$ where perturbation theory applies. We call this integral
$I(\rho)$ and get that $F_{IR}=\lim_{\rho \rightarrow \infty}I(\rho)$.

$I(\rho)$ is shown to be the Borel transform of $F(s)$ of order $k$, where
$k$ is related to the way we analytically continue in the complex scale
parameter. 
The Borel transform has also been applied to perturbation theory
in \cite{LGZJ} where the Borel transform in the coupling is used. 
In this case assumptions (justified by details of the dynamics) 
have to be made about the poles of the Borel transform.
The essential difference compared with 
our approach is that we transform with respect
to the scale parameter. The analytic properties of Green functions,
which are needed to define the Borel transform, in this case 
follow directly from general principles of quantum field theory.   

We will use this method on two examples where the infra-red limit of the
physical quantity $F(s)$ in question 
is well known, namely the central charge of the 
unitary minimal models perturbed by the relevant field
$\phi_{(1,3)}$, and the critical exponents of $\varphi^4$ theory in three
dimensions. 

We also considered the perturbed minimal models in \cite{lkpm} where this
method was shown to improve upon renormalisation group  improved
perturbation theory by a factor 2. 
$F(s)$ is given as the two point function of the
energy momentum tensor, and we will here  address some of the details
left out of our discussion in  \cite{lkpm}. 
We first show the analytic properties of $F(s)$, and we
then obtain a bound on the exact non perturbative expression for 
$I(\rho)$, and this  
allows us to obtain an approximate value for
calculating the central charge. 

We employ this approximation method to calculate the central charge for
the free fermionic and bosonic theories perturbed by a mass term and the
unitary minimal models perturbed by the relevant operator $\phi_{(1,3)}$.
The free massive theories can be solved exactly and these then demonstrate 
how $I(\rho)$ behaves in an exact case. For the perturbed minimal models we
first calculate $F(s)$ as a renormalisation group improvement of a
one loop perturbative calculation and show that this expression has the
correct limit when $m \rightarrow \infty$, where $m$ characterises the
minimal model. We then show how our method improves upon the
perturbative renormalisation group result.  If we use the largest domain of
analyticity for $F(s)$ apparent from the spectral decomposition there
is something of a surprise. The approximation becomes
dramatically
close to the exact value, which is remarkable as it is
only based on one loop perturbation theory. 

The other example is $\varphi^4$ theory in three dimensions, where we
calculate the critical exponents $\nu$ and $\eta$. $\varphi^4$ theory
has
been shown to be Borel summable of order one in the perturbative expansion in
the coupling constant. The perturbative 
series will often at most be asymptotic, if for example a
theory is ill defined for negative couplings the convergence radius will be
zero. 
The question is then if the perturbative series captures all of the
physics when it is only asymptotic. This will depend on the analytic
properties of the correlators, and if the theory is Borel summable perturbation
theory will contain all information about the theory.
We  use the propagator to define functions
$F_{\nu}, F_{\eta}$ which have the critical exponents $\nu$ and $\eta$ 
as their infra-red limit. All scale dependence is moved into the coupling using the
running coupling so that $F_{\nu}(s)$ and $F_{\eta}(s)$ 
are given as asymptotic series
in the scale parameter $s$, and we show their analyticity. 
In this example the actual calculation of the physical quantities $\nu$ and
$\eta$ is different form the minimal models case,  as we cannot
use
the same approximation. The
reason  is that in this case we use higher order perturbative expressions
(5,6 and 7 loops) and one cannot find an exact expression for the
perturbative running coupling (unlike the one loop case). The infra-red
values for $\nu, \eta$ are therefore obtained from the limiting value of
the Borel transform, and to calculate this we use a conformal mapping and
Pad\'{e} approximation which is done in the appendix. The estimates we obtain
for $\nu, \eta$ are within the errors of other more elaborate 
approximations.

The main results are that we can use analyticity in the scale parameter to
express the infra-red physics as an integral in the ultra-violet region, the infra-red limit is then
obtained taking the limit in the Borel transformed quantity.  
We show how this works in two examples with well known infra-red limits.

The outline is as follows: In the next section we will 
discuss analyticity of correlators, 
introduce the contour integral
giving the infra-red limit and show that the contribution from the cut vanishes for
large $\rho$. We also show that the contour integral is the Borel
transform of
$F(s)$. In section 3 we define $F(s)$ for the central charge, show 
its analyticity and discuss an approximation method for calculating the
central charge. An exact bound on $I(\rho)$ is given. In section 4
we use this approximation on the free massive bosonic and fermionic theory
together with the minimal models perturbed by $\phi_{(1,3)}$. 
In section
5 we discuss the $\varphi^4$ theory and introduce a function giving the
critical exponents $\nu, \eta$ in the infra-red
 limit, and obtain estimates of
$\nu$ and $\eta$  from the limit of the Borel transform.


\section{Defining the contour integral}
The physical quantities of a quantum field theory can be written in terms of Green functions which
have well known analyticity properties. As an example consider the two point
function of a scalar field $A(p)$. The K\"{a}llen-Lehmann spectral 
representation writes this as a sum over intermediate states
\begin{equation}
  \label{kallenlehmann}
  \langle A(p)A(0)\rangle = \int_0^\infty\! d\mu^2\ \tilde{c}(\mu^2)\frac{1}{p^2+\mu^2}
\nonumber
\end{equation}
where $d\mu^2 \tilde{c}(\mu^2)$ is the spectral density. 
Introducing the complex scale factor $s\in \mathbb{C}$
\begin{equation}
  \label{scalefac}
  \langle A(sp)A(0)\rangle = \int_0^\infty\! d\mu^2\ \tilde{c}(\mu^2)\frac{1}{s^2p^2+\mu^2}
\nonumber
\end{equation}
then shows that $\langle A(sp)A(0)\rangle$ is analytic in the positive 
half-plane $\Re (s)>0$. 

Generally physical quantities will not be analytic in the positive half-plane
but in some sector $S(\alpha)$ of the complex plane, defined as 
\begin{equation}
\label{sector}
S(\alpha)=\ \{z=re^{i\phi}\ |\
0<r<\infty,\ -\tfrac{\alpha}{2}
<\phi<\tfrac{\alpha}{2} \}.
\end{equation}
Note that $0\notin S(\alpha)$.
We want to calculate the infra-red limit $F_{IR}$ of a physical quantity $f(x)$,
where $F_{IR}=\lim_{|x|\rightarrow \infty}f(x)$, and we will denote the UV
limit $F_{UV}=\lim_{|x|\rightarrow 0}f(x)$. We are considering functions
which have well defined limits in $S(\alpha)$, i.e.\ 
$\lim_{|s|\rightarrow \infty}F(s)=F_{IR}$ 
and $\lim_{s\rightarrow 0}F(s)=F_{UV}$ for $s\in S(\alpha)$.

In the two examples we will consider the physical quantities $f(x)$
are given
as functionals of a two point correlator resulting in analytic functions
$F(s)$. A large class of physical quantities will in this way preserve the
analytic structure of the Green functions in the theory. 

The scale parameter $s$ can of course be introduced into the physical
quantity in a number of ways. We will define by $\tilde{F}(s)$ the analytical
continuation of $f(x)$ where $s$ is introduced as $\tilde{F}(s)\equiv
f(sx)|_{x=1}$ so that $\tilde{F}(s)$ has the expansion $\tilde{F}(s)=\sum
\tilde{F}_n s^n$ around the origin, and $\tilde{F}(s)$ is analytic in
$S(\alpha)$. 
Choosing another positive power
$f(s^\gamma x)|_{x=1}$, $\gamma>0$, gives the same ultra-violet and infra-red limits, but
different intermediate behaviour. 
Let us now
introduce the scale parameter so that the opening of the analytic sector is
$2(\pi-\epsilon')$ with $\epsilon'\ll 1$. $F(s)=\tilde{F}(s^a)$ is analytic in
$S=S(2(\pi-\epsilon'))$ provided that
$a=\tfrac{\alpha}{2(\pi-\epsilon')}$.   
We will now express the infra-red limit as an integral entirely in the ultra-violet region using
analyticity. 

Using Cauchy's theorem the analyticity of $F(s)$ implies that the contour 
integral
\begin{equation}
  \label{nycontourint2}
  \frac{1}{2(\pi-\epsilon) i}\int_C ds \, \frac{e^{\rho /s}}{s}F(s)=
\frac{1}{2(\pi-\epsilon) i}\left(\int_{C_0}ds+\int_{C_1}ds+\int_{C_2}ds\right)\frac{e^{\rho /s}}{s}F(s)=0,
\nonumber
\end{equation}
vanishes in the sector $S$, where $\rho \in \mathbb{R}_+$ and 
the contour $C$ is given in figure 1;  
$\epsilon>\epsilon'$ so that $C\subset S$. 
In the limit
where $r_{IR}\rightarrow \infty$,
the contribution from the contour $C_{2}$  (see figure
1)  becomes
\begin{equation}
  \label{nyC22}
\lim_{r_{IR} \rightarrow \infty}
\frac{1}{2(\pi-\epsilon) }\int_{\pi-\epsilon}^{-\pi+\epsilon}d\theta \, e^{\rho e^{-i\theta}/r_{IR}}
F(r_{IR}e^{i\theta})= -\lim_{|s|\rightarrow \infty}
F(s) = -F_{IR}.
\end{equation}
The angular integral and the limit $r_{IR}\rightarrow \infty$ can be
interchanged as the integrand in \eqref{nyC22} is bounded by a constant
in the limit $r_{IR}\rightarrow \infty$.
We then get the integral representation of $F_{IR}$
\begin{equation}
F_{IR}=\frac{1}{2(\pi-\epsilon) i}\left( \int_{C_0}ds \frac{e^{\rho /s}}{s}F(s)+
\int_{C_1}ds \frac{e^{\rho /s}}{s}F(s)  \right) .
\label{nyCir}
\end{equation} 
Analogously by considering 
 \begin{equation}
  \label{nycontourint}
  \frac{1}{2(\pi-\epsilon) i}\int_C ds \, \frac{e^{\rho
      s}}{s}F(s)=\frac{1}{2(\pi-\epsilon)  i}
\left(\int_{C_0}ds+\int_{C_1}ds+\int_{C_2}ds\right)\frac{e^{\rho s}}{s}F(s)=0,
\nonumber
\end{equation}
we get that 
\begin{equation}
F_{UV}=\frac{-1}{2(\pi-\epsilon) i}\left(\int_{C_2}ds \frac{e^{\rho s}}{s}F(s)+
\int_{C_1}ds \frac{e^{\rho s}}{s}F(s)  \right).
\label{nycuv}
\end{equation}
We denote the integral along the contour $C_1$ (close to the cut) by
\begin{equation}
cut(\rho)=\frac{1}{2(\pi-\epsilon) i}\int_{C_1}ds \frac{e^{\rho/ s}}{s}F(s).
\end{equation} 
The integrand in this integral is damped by the factor $e^{-\rho/ |s|}$, and
we will show that $\lim_{\rho \rightarrow \infty}cut(\rho)=0$ by showing that
it is bounded by a finite integral for all $\rho$ allowing us
to take the limit $\rho \rightarrow\infty$ in the integrand.
We substitute $s=re^{\pm i(\pi-\epsilon)}$ for points on $C_1$ in the upper
and lower half-plane, the integral becomes
\begin{equation}
  \label{label1}
\begin{split}
  cut(\rho)=&\ \frac{-1}{2(\pi-\epsilon) i}\left(\int_{r_{IR}}^{r_{UV}}dr \frac{e^{\rho 
      e^{-i(\pi
        -\epsilon)}/r}}{r}F(re^{i(\pi-\epsilon)}) +\! \int_{r_{UV}}^{r_{IR}}dr\frac{e^{\rho 
      e^{i(\pi -\epsilon)}/r}}{r}F(re^{-i(\pi-\epsilon)}) \right)\\
=&\
\frac{-1}{(\pi-\epsilon)}\int_{r_{UV}}^{r_{IR}}dr \frac{e^{\rho \cos (\pi
    -\epsilon)/r}}{r}\mbox{Im} [ e^{i\rho \sin (\pi-\epsilon)/r}F(re^{-i(\pi-\epsilon)}) ]
      \\
=&\ \frac{-1}{(\pi-\epsilon)}\int_{r_{UV}}^{r_{IR}}dr\frac{e^{\rho \cos (\pi
    -\epsilon)/r}}{r}\left(\cos (\rho \sin
      (\pi-\epsilon)/r)\mbox{Im}[F(re^{-i(\pi-\epsilon)})] \right. \\ 
& \left. \ \ \ \ + \ \ \mbox{Re}[F(re^{-i(\pi-\epsilon)})]
\sin (\rho \sin (\pi-\epsilon))\right).
\end{split}
\nonumber
\end{equation}
We divide the $r$ interval into $( r_{UV},1)$ and $( 1,r_{IR})$ and write
$cut(\rho)=cut_{UV}+cut_{IR}$. The function $F(s)$ is finite in $S$ hence
there exists a constant $q>0$ so that in the limit $r_{UV}\rightarrow 0$ 
\begin{equation}
  \label{iir1}
  |cut_{UV}|< q \int_0^1 dr \frac{e^{-\rho /r}}{r}\rightarrow 0 \ \ 
\mbox{    for } \ \rho \rightarrow \infty.
\end{equation}
We know that $F(s)\rightarrow F_{IR}\in \mathbb{R}$ for $|s|\rightarrow
\infty$ from $S$, hence $\mbox{Im}[F(re^{-i(\pi-\epsilon)})]\rightarrow 0$ for
$r\rightarrow \infty$. If $\mbox{Im}[F]$ falls off like $r^{-\delta}$ for
some $\delta>0$ then in the limit $r_{IR}\rightarrow \infty$, $\exists\ k_1,k_2,k_3>0$: 
\begin{equation}
  \label{cutirlim}
  \left|cut_{IR}\right|<k_1\int_1^\infty dr \frac{\left(\sin
  (k_2/r)+k_3r^{-\delta}\right)}{r}<\infty \mbox{   for all  } \rho \in
  \mathbb{R}_+ . 
\end{equation}
For a general $F(s)$ where the fall off might be slower we keep a finite
$r_{IR}$ then $|cut_{IR}|$ is again finite for all $\rho$ and a finite
$r_{IR}$ introduces a $O(\tfrac{1}{r_{IR}})$ term in \eqref{nyCir} which is
negligible for $r_{IR}$ large. As $\epsilon\ll 1$ we can replace
$\tfrac{1}{2(\pi-\epsilon)i}$ with $\tfrac{1}{2\pi i}$ given a term
$O(\epsilon)$ on the right hand side in \eqref{nyCir} which is again 
negligible for small $\epsilon$.
We can then define 
\begin{equation}
  \label{firrho}
  I_{IR}(\rho)=\lim_{r_{UV}\rightarrow 0}\ 
\frac{1}{2\pi i}\int_{C_0} ds \frac{e^{\rho/s}}{s}F(s)
\end{equation}
so that $F_{IR}=\lim_{\rho \rightarrow \infty}I_{IR}(\rho)$ and we have 
succeeded
in writing the infra-red limit as the limiting value of a contour integral 
in the ultra-violet region. 

We will now  
use Cauchy's theorem again to rewrite $I_{IR}(\rho)$ and this will show that
$I_{IR}(\rho)$ is the Borel transform of $\tilde{F}(s)$. 
Instead of integrating over $C_0$ we will integrate over $\tilde{C}$ given as
the path from the origin along the ray $\arg (s)=-\pi+\epsilon$ 
and then anti-clockwise along
$|s|=\tilde{x}>0$ until $\arg (s)=\pi-\epsilon$ and then back to the origin. 
It follows that this contour integral is independent of the choice of 
$\tilde{x}$ and $\epsilon$ as long as $\epsilon'<\epsilon<\tfrac{\pi}{2}$. 
The upper limit ensures that 
the rays stay in the negative half-plane (where $e^{\rho/s}$ is a damping
factor), and the lower limit that 
$F(s)$ is analytic on the contour. 
Using this contour the ultra-violet limit  
$r_{UV}\rightarrow 0$ can be taken explicitly by extending the rays to the
origin.

Using that $F(s)=\tilde{F}(s^a)$ for $a=\tfrac{\alpha}{2(\pi-\epsilon')}$
then amounts to 
\begin{equation}
  \label{newcontour}
  I_{IR}(\rho)=\frac{1}{2\pi i}\int_{\tilde{C}}\frac{ds}{s}e^{\rho/s}F(s)=
\frac{1}{2\pi i}\int_{\tilde{C}}\frac{ds}{s}e^{\rho/s}\tilde{F}(s^a)=
\frac{k}{2\pi i}\int_{\tilde{C}'}\frac{ds}{s}e^{(\frac{\tilde{\rho}}{s})^k}
\tilde{F}(s),
\end{equation}
where $k=\tfrac{1}{a}$, $\tilde{\rho}=\rho^{1/k}$ and 
$\tilde{C}'$ is the contour where the rays satisfy
$|\arg(s)|=|(\pi-\epsilon'')/k|$ for any $\epsilon'' \in
(\epsilon',\tfrac{\pi}{2})$.  
We will write $\epsilon''=(\pi-\tilde{\epsilon})/2$ for
any $\tilde{\epsilon}\in (0,\pi-2\epsilon')$, 
then $|\arg((\tfrac{\tilde{\rho}}{s})^k)|>\pi/2$
and the integrand in \eqref{newcontour} is again damped on the rays. 
Also, on the rays is $|\arg
(s)|=(\pi+\tilde{\epsilon})/2k=\tfrac{\alpha}{4}\tfrac{\pi+\tilde{\epsilon}}{\pi-\epsilon'}<\alpha/2$
so that $\tilde{F}(s)$ is analytic on the contour, $\tilde{C}'\subset
S(\alpha)$, and the contour integral is therefore well defined.

Equation \eqref{newcontour} then shows that $I_{IR}(\tilde{\rho})$ is the Borel
transform of $\tilde{F}(s)$ of order $k$ \cite{balser}. 
The Borel transform of order $k$ of a formal power series $h(z)=\sum_n
h_nz^n$ is given by ${\cal B}_k (h)(\rho)=\sum_n \tfrac{h_n\rho^n}{\Gamma
  (1+n/k)}$, using the integral representation $\tfrac{1}{\Gamma
  (1+n)}=\tfrac{1}{2\pi i}\int_{\tilde{C}'}\tfrac{ds}{s}e^{1/s}s^n$ it then
follows that \eqref{newcontour} is the Borel transform of order $k$ of
$\tilde{F}(s)$ by inserting a series expansion for $\tilde{F}(s)$.

The contour integral in \eqref{newcontour} is independent of the contour
$\tilde{C}'$, i.e.\ in 
the choice of $\tilde{\epsilon}$ for $\tilde{\epsilon}\in
(0,\pi-2\epsilon')$. This shows that equation \eqref{newcontour} holds for
all $k$ with $F(s) \equiv \tilde{F}(s^{1/k})$ and $\alpha>\pi/k$, because we can
always find an $\tilde{\epsilon}>0$ so that
$\alpha>(\pi+\tilde{\epsilon})/k$, which again implies that $\tilde{C}'\subset
S(\alpha)$ and $|\arg (\tilde{\rho}/s)^k| > \pi/2$ which again 
makes the contour integral well defined. 

We have then shown that: {\it the infra-red limit of a physical quantity, $F_{IR}$, is
the limiting value of the Borel transform of $\tilde{F}(s)$, and changing the
way in which the scale parameter 
is introduced amounts to changing the order of the
Borel transform. The order $k$ has to satisfy the bound $\alpha>\pi/k$ where
$\alpha$ determines the analytic sector 
of the physical quantity $\tilde{F}(s)$}.

One way of calculating the contour integral in \eqref{newcontour} is to
insert a series expansion of $\tilde{F}(s)$ around the origin, but because
$\tilde{F}(s)$ is only analytic in a sector this series can only be an
asymptotic expansion of $\tilde{F}(s)$
\footnote{If $f(z)$ is an analytic function in a neighbourhood
of the origin it follows that its asymptotic series around the origin 
is convergent and equal to the power series of $f(z)$; 
the lack of analyticity at the origin forces the asymptotic series to
diverge. A formal power series $f(z)=\sum f_n z^n$ is asymptotic of order $k$
if $\exists\ C,K>0: |f_n|< C K^n \Gamma (1+n/k)$, this shows that the Borel
transform of 
order $k$, of such a series, has a non zero convergence radius.}.
An analytic function has a unique
asymptotic expansion, but an asymptotic series might be the asymptotic
expansion of several analytic 
functions.
However in the case where a function is analytic in a sector of opening
$\alpha>\tfrac{\pi}{k}$, which is the situation we have above, 
the map between the analytic function and the
asymptotic series of order $k$ is injective (but not surjective)
\cite{balser}. This means
that all information about the exact function is contained in the
asymptotic series. 

An asymptotic series of $\tilde{F}(s)$ in the scale parameter can be obtained
doing the perturbative expansion in the coupling and then inserting the
running coupling constant.
Assume that $\tilde{F}$ is given as a formal perturbative series, which we
write as  $\tilde{F}(s,g)= \sum_n \hat{F}_n(s) g^n$. 
The Callan-Symanzik equation states that
a theory is invariant, i.e.\ the correlators are invariant,  
under a scale transformation if the couplings change according to the 
renormalisation group. 
A scaled quantum field theory can therefore equivalently be described by a 
theory on the same
scale, but with couplings changed according to the renormalisation group.
In this way
scale dependence of a theory can be moved into the running coupling 
$\bar{g}(s)$.
Moving all scale dependence into the running coupling we get an asymptotic
series in the scale 
$\tilde{F}(s=1,\bar{g}(s))=\sum \tilde{F}_n s^n\ $ \footnote{If $\tilde{F}(s)$ 
is asymptotic of order
$k'$ ($\tilde{F}(s)\in \mathbb{C}[[s]]_{1/k'}$) we will choose $k=k'$, and then
the
Borel transform of $\tilde{F}$ has a
non-zero convergence radius and the analytically continued value at infinity
uniquely determines $F_{IR}$. The analytical continuation can be done
  by a conformal mapping, we will discuss this in the appendix.}.

The examples we consider below
have $\alpha=\pi-\epsilon'$, where $\epsilon'\ll 1$, hence from the 
constraint $\alpha>\tfrac{\pi}{k}$: $k=1+\delta$ for some
$\delta>0$. We will generally try to minimise the order $k$ ($k\rightarrow
1$), thus maximising the analytic sector and the convergence of the Borel 
transform.


\section{Defining $\tilde{F}(s)$ for the central charge}
As the first example we want to calculate the infra-red central charge of a 2
dimensional
unitary quantum field theory. The renormalisation group 
flow of the central charge is governed 
by Zamolodchikov's
$c$-theorem \cite{zamolodchikov}. 
This states that for a unitary and
renormalisable quantum field theory in 2 dimensions  there exists a function which is
monotonically decreasing along the renormalisation group flow, and which is
stationary only for conformally invariant theories where it takes the value of
the Virasoro central charge.
The $c$-theorem implies that the infra-red limit, where the scale goes to
infinity, and the ultra-violet  limit, where the scale vanishes, are
fixed points of the renormalisation group. 
In 2 dimensions scale invariance implies conformal invariance so in these scaling 
limits we have a conformal field theory characterised by the central
charge. We define the energy-momentum tensor as in \cite{cappelli}
\begin{equation}
  \label{Tdef}
  \langle T_{\mu \nu}(x)\rangle = \frac{2V}{\sqrt{g(x)}}
  \frac{\delta W[g]}{\delta g^{\mu\nu}},
\end{equation}
where $V=\mbox{Vol}(S^{n-1})=2 \pi$ in 2D and $W[g]$ is the effective action.

The correlator $\langle T_{zz}(z)T_{zz}(0)\rangle$ gives the ultra-violet and
infra-red central charges in the limits $z \rightarrow 0$ and 
$z \rightarrow \infty$ respectively. In \cite{cappelli} 
$\langle T_{\mu \nu}(x)T_{\rho \sigma}(0)\rangle$ was written using the
K\"{a}llen-Lehmann  spectral representation:
\begin{equation}
  \label{correllator}
\langle T_{\mu \nu}(x)T_{\rho \sigma}(0)\rangle = 
\frac{\pi}{3\cdot 16}\int_0^\infty  d\mu^2 \, \tilde{c}(\mu^2)
\int \frac{d^2 p}{(2\pi)^2} e^{ipx}\, \frac{(g_{\mu \nu}p^2-p_\mu
  p_\nu)(g_{\rho \sigma}p^2 - p_\rho p_\sigma)}{p^2+\mu^2} ,
\nonumber
\end{equation}
and it follows that
\begin{equation}
  \label{spectral}
\langle T_{zz}(z,\bar{z})T_{zz}(0,0)\rangle = 
\frac{\pi}{3\cdot 16}\int_0^\infty  d\mu^2 \, \tilde{c}(\mu^2)
\int \frac{d^2 p}{(2\pi)^2} \frac{e^{\frac{i}{2}(p\bar{z}+\bar{p}z)}}
{p\bar{p}+\mu^2} \, \bar{p}^4,
\end{equation}
where we use the usual complex variables\footnote{$z=t_E+ix,
  \bar{z}=t_E-ix$ and $d^2z \equiv d^2x=dx \wedge dt_E=-\tfrac{i}{2} 
dz\wedge d\bar{z}$.}  $z,\bar{z}$, and
$\tilde{c}(\mu^2)d\mu^2$ is the spectral density which represents the density
in degrees of freedom of the quantum field theory at the mass $\mu$. 
If we scale $z,\bar{z}$ by a positive real dimensionless parameter $s$
($s>0$) we  get
\begin{equation}
  \label{Fscaled}
\langle T_{zz}(s z,s\bar{z})T_{zz}(0,0)\rangle=
\frac{\pi}{3\cdot 16}\int_0^\infty d\mu^2 \, \tilde{c}(\mu^2)
\int \frac{d^2 q}{(2\pi)^2} \frac{e^{\frac{i}{2}(q\bar{z}+\bar{q}z)}}
{q\bar{q}+s^2\mu^2} \, \frac{\bar{q}^4}{s^4}.
\end{equation}
In the ultra-violet limit where $s\rightarrow 0$ then 
$\frac{\bar{q}^4}{q\bar{q}+s^2\mu^2}\rightarrow \frac{\bar{q}^4}{q\bar{q}}$
and \eqref{Fscaled} becomes\footnote{The Fourier transform of $(\pi/24)
  \bar{p}^4 /\bar{p}p$ is $1/z^4$.}
\begin{equation}
  \label{ultrav}
\langle  T_{zz}(sz,s\bar{z})T_{zz}(0,0)\rangle \rightarrow 
  \frac{1}{2s^4z^4}\int_0^\infty d\mu^2 \, \tilde{c}(\mu^2)=
  \frac{c_{UV}}{2s^4z^4} \ \ \mbox{for} \ s \rightarrow 0.
\end{equation}
The ultra-violet central charge is therefore $c_{UV}=\int_0^\infty d\mu \, \tilde{c}(\mu)$. 
To calculate the infra-red limit we first note that 
\begin{equation}
  \label{frie}
 \int \frac{d^2 q}{(2\pi)^2} \frac{e^{\frac{i}{2}(q\bar{z}+\bar{q}z)}}
{q\bar{q}+s^2\mu^2} \, \bar{q}^4 = 2^4 \left(\frac{\p}{\p z}\right)^4 
G(z,\bar{z},s\mu),
\end{equation}
where $G(z,\bar{z},\mu)$ is equal to the free Bose propagator at mass $\mu$. 
$G(z,\bar{z},\mu)$ can be written in terms of a modified Bessel function \cite{francesco}
\begin{equation}
  \label{boseG}
  G(x,\mu)=\int \frac{d^2p}{(2\pi)^2}\frac{e^{i x\cdot p}}{p^2+\mu ^2}=
\frac{K_0(|x|\mu)}{2\pi}.
\end{equation}
Performing the differentiation we can then write \eqref{Fscaled} as
\begin{multline}
\langle  T_{zz}(sz,s\bar{z})T_{zz}(0,0)\rangle = 
\frac{1}{2\cdot 48\, s^4 z^4}
\int_0^{\infty}d\mu^2\ \tilde{c}(\mu^2)\mu s|z|((\mu^3 s^3
  |z|^3  \\ + 24\mu s|z|)K_0(\mu s|z|)   
  +(8\mu^2 s^2 |z|^2+48)K_1(\mu s|z| )) .
\label{ftilde1}
\end{multline}
In the infra-red limit where $s\rightarrow \infty\ $ $K_0(\mu |z|s)$
and $K_1(\mu |z|s)$ have the asymptotic
behaviour \cite{stegun} $e^{-\mu s | z |}$ and the only
contribution to \eqref{ftilde1} comes from the massless limit where $\mu
\rightarrow 0$, hence
\begin{equation}
  \label{infrar}
\langle  T_{zz}(sz,s\bar{z})T_{zz}(0,0)\rangle  \rightarrow 
\lim_{\epsilon \rightarrow 0}\, \frac{1}{2s^4z^4}
\int_0^\epsilon d\mu^2 \, \tilde{c}(\mu^2)=\frac{c_{IR}}{2s^4z^4}\ \ \ \mbox{for
  }s\rightarrow \infty,
\end{equation}
so that $c_{IR}=\lim_{\epsilon \rightarrow 0}\int_0^\epsilon d\mu^2 \, \tilde{c}(\mu^2)$.
This shows that $c_{UV}\ge c_{IR}$ as the spectral density is positive for a
unitary theory, and this is another way of showing the
$c$-theorem \cite{cappelli}. This representation of the central charge using
the spectral representation also shows that the central charge measures the
number of (massless) degrees of freedom of the CFT. We define the 
function\footnote{When we set $|z|=1$ then $\mu$ becomes dimensionless in
  \eqref{ftilde1}, we also denote this dimensionless quantity by $\mu$.} 
\begin{equation}
\left.
\tilde{F}(s)=2z^4s^4\langle
T_{zz}(s z,s \bar{z})T_{zz}(0,0)\rangle\right|_{z=\bar{z}=1}
\end{equation} 
here $s$ and $\tilde{F}(s)$ are dimensionless and $s\in \mathbb{R}_+$. 
This function then satisfies 
\begin{equation}
\label{Fgraense}
\tilde{F}(s)\rightarrow \left\{ 
\begin{array}{ll}
c_{UV} & \mbox{for }s\rightarrow 0_{+}, \\
c_{IR} & \mbox{for }s\rightarrow \infty. 
\end{array} \right.
\end{equation}

\subsection{Analyticity of $\tilde{F}(s)$}
We will show that $\tilde{F}(s)$ is an analytic continuation 
of $\tilde{F}(x),\ x\in \mathbb{R}_+$, for 
$s\in S=S(\pi-\epsilon')$ (with $\epsilon'\ll 1$); 
to show this write 
$\tilde{F}(s)=\int_M d\nu f(s,\nu)$. For $\tilde{F}$ 
to be holomorphic in $S$ 
then $f(s,\nu)$ must be holomorphic in $S$ for all $\nu \in M$, and both $f$ 
and $\frac{df}{ds}$ must be integrable over the set $M$. 
Using (\ref{ftilde1}) above we can write 
$\tilde{F}(s)$ as
\begin{equation}
  \label{F2}
\tilde{F}(s)=\frac{1}{48}\int_0^\infty\! \! \! d\mu^2 \, \tilde{c}(\mu^2)
 \mu s \left( ({\mu}^{3}{s}^{3}\!+\!24\mu 
s) K_0(\mu s)\!+\!(8 \mu^2 s^{2}\!
+\!48) K_1(\mu s)\right ).
\end{equation}
$K_{\nu}(z)$ is holomorphic in $S$ so $f(s,\nu)$ is clearly holomorphic in
$S$.  The modified Bessel functions also satisfy that $|K_\nu(z)|$ is bounded
for  $|z|\geq \epsilon$ for any $\epsilon\in \mathbb{R}_+$ and $|\arg z|<\frac{\pi}{2}$,
which is the case for $z=\mu s$ 
when $s\in S$. For large values of $|s|$
integrability is ensured by $\int_0^{\infty}d\mu^2\
\tilde{c}(\mu^2)=c_{UV}<\infty$ and the asymptotic behaviour $K_\nu(z)=
\sqrt{\frac{\pi}{2z}}e^{-z}(1+O(z^{-1}))$. Around the origin $K_0(z) z\rightarrow 0$ for $z\rightarrow 0$ and
$K_1(z) z\rightarrow 1$ for $z\rightarrow 0$, hence $\tilde{F}(s)$ 
is integrable. 
$\frac{d \tilde{F}(s)}{ds}$ 
is shown to be integrable in a similar way and the 
analyticity is shown. 

From the form of $\tilde{F}(s)$ in (\ref{F2}) it follows that $\tilde{F}(s)$
 has a limit value $\tilde{F}(s)\rightarrow d$ 
for $s\rightarrow 0$ from $S$. From
(\ref{Fgraense}) it follows that $d=c_{UV}$, and we define 
$\tilde{F}(0)=c_{UV}$. 
It also follows from (\ref{F2}) that $\tilde{F}(s)$
has a limit value for $|s|\rightarrow \infty$ from $S$:
$\tilde{F}(s)\rightarrow d'$ and (\ref{Fgraense}) again sets $d'=c_{IR}$. 


\subsection{The approximation for $c_{IR}$}
We have the following representation of  $c_{IR}$ and 
$c_{UV}$
\begin{align}
c_{IR} &=\frac{1}{2\pi i}\left( \int_{C_0}ds \frac{e^{\rho /s}}{s}F(s)+
\int_{C_1}ds \frac{e^{\rho /s}}{s}F(s)  \right)=I_{IR}(\rho)+cut(\rho)
\label{align2} \\
c_{UV} &=\frac{-1}{2\pi i}\left(\int_{C_2}ds \frac{e^{\rho s}}{s}F(s)+
\int_{C_1}ds \frac{e^{\rho s}}{s}F(s)  \right)=I_{UV}(\rho)+cut(\rho), \label{align1} 
\end{align}
where in both cases the contribution from the cut is rapidly decreasing in
$\rho$. Note that in these relations the integral is performed
in the opposite scaling limit of the quantity we calculate. 
We will here concentrate on \eqref{align2} as perturbation theory can be
applied to $F(s)$ when $s\in C_0$, we will call $I_{IR}(\rho)=I(\rho)$.

We can choose coordinates in the coupling constant space so
that the ultra-violet fixed point corresponds to $\bar{g}(s)=0$. In the ultra-violet limit where
$s\rightarrow 0$ we may describe $F(s)$ by perturbation theory as
$\bar{g}(s)\rightarrow 0$. The nth order perturbative approximation of
$F(s)$ is denoted by $F_n(s)$ and the corresponding integral by 
$I_n(\rho)$. 
In the limit of large $\rho$ the contribution from the cut vanishes and we
get 
\begin{equation}
  \label{cirgranse}
  c_{IR}\approx \lim_{\rho \rightarrow \infty}I_n(\rho) =\lim_{s\rightarrow 
\infty}F_n(s)
\end{equation}
where the last equality follows setting $s'=s/ \rho$ in \eqref{align2} and
then taking the limit $\rho \rightarrow \infty$ in the integrand valid for
all $r_{UV}>0$.
Moving all scale dependence into the running coupling $\bar{g}(s)$ we can
write $F_n(s)=\Phi_n(\bar{g}(s))$. Let $g^*_{IR}$ denote the first non
trivial zero of the perturbative $\beta-$function, i.e.\
$\bar{g}(s)\rightarrow g^*_{IR}$ for $s\rightarrow \infty$. Then equation
\eqref{cirgranse} becomes
\begin{equation}
  \label{graense1}
  \lim_{s\rightarrow \infty}F_n(s)=\lim_{s\rightarrow
  \infty}\Phi_n(\bar{g}(s))=\Phi_n(g^*_{IR})=c^*_{IR}
\end{equation}
which is the perturbative estimate of $c_{IR}$ we want to improve. 

$F_n(s)$ is the ultra-violet perturbative approximation to $F(s)$ 
and the integration range in $I(\rho)$ is compact so 
$\lim_{\rho \rightarrow 0}I_{n}(\rho)=\lim_{\rho \rightarrow
  0}I(\rho)=c_{UV}$.  
$I_n(\rho)$ therefore provides a good approximation to $I(\rho)$,
for small enough $\rho$,
since the power series expansion of $I(\rho)$ is controlled by
the small $s$ expansion of $F(s)$ for which perturbation theory applies. 
This
is illustrated in \eqref{gammaapprox} below for the minimal models.
For larger values of $\rho$ higher order terms in the expansion of 
$I_n(\rho)$ become important and the coefficients of the expansion of 
$I_n(\rho)$ and $F_n(\rho)$ part company. 
If $c_{IR}<c^*_{IR}$
and if the region where $I_n(\rho)$ 
is a good approximation to $I(\rho)$ is large enough, then
$I_n(\rho)$ will have a minimum before approaching
its limiting value of $c^*_{IR}$. 
Since this minimum occurs at the largest value of $\rho$ for which $I_n(\rho)$
is a reasonable approximation to $I(\rho)$ and the true value of
$c_{IR}$ is given by $I(\infty)$, it is this minimum of $I_n(\rho)$
that we will use to provide a better estimate of $c_{IR}$.
The approximation then becomes
\begin{equation}
  \label{approxcir}
  c_{IR}=I_n(\rho_m)
\end{equation}
where $\rho_m$ is the value where $I_n(\rho)$ attains its minimum.
Below we consider the quantity $\Delta c_{exact}=c_{UV}-c_{IR}$ and denote
the approximation to it $\Delta c_{approx}=c_{UV}-I_{n}(\rho_m)$, we call the
perturbative value $\Delta c_{pert}=c_{UV}-\lim_{\rho\rightarrow
  \infty}I_n(\rho)=c_{UV}-c_{IR}^*$. 
This approximation rests upon the assumption that the
exact function is monotonically decreasing from $c_{UV}$ to $c_{IR}$, or at
least that its minimum value is close to $c_{IR}$.  Below we will show that
this is indeed the case for $k=4$, at least to a very good approximation.


\subsection{Exact bound on $\Delta c$}
As discussed above can the exact value $I(\rho)$ possibly 
be smaller than the asymptotic value
$c_{IR}$. We will denote by ${\rho'}_m$ the value where $I(\rho)$ attains 
its minimum and 
$\Delta c_{est}=c_{UV}-I({\rho'}_m)$, then 
$\Delta c_{exact}-\Delta
c_{est}$ measures the undershoot of the exact function $I(\rho)$ compared with
its asymptotic value $c_{IR}$ (see figure 3). 
From the spectral representation of $F(s)$ in \eqref{F2} we can obtain a
rigorous lower bound on this exact undershoot.
In \cite{cappelli} it was shown that the spectral density can be written as
\begin{equation}
\nonumber
\tilde{c}(\mu^2)=c_{IR}\delta(\mu^2)+\hat{c}(\mu^2),
\end{equation}
we showed in \eqref{ultrav} that $\int d\mu^2 \tilde{c}(\mu^2)=c_{UV}$ hence
$\int d\mu^2 \hat{c}(\mu^2)=\Delta c_{exact}$.
Using this in \eqref{F2} we get
\begin{equation}
  \label{estimate}
\begin{split}
  \Delta c_{est}=\ & c_{UV}-I({\rho'}_m)=c_{UV}-\frac{1}{2\pi
  i}\int_{\tilde{C}}\frac{ds}{s}e^{{\rho'}_m/s}F(s) \\  =\ &
\Delta c_{exact}-\int d\mu^2\
  \hat{c}(\mu^2)\Upsilon(\mu^2, {\rho'}_m) \nonumber
\end{split}
\end{equation}
where
\begin{equation}
  \label{psi}
\begin{split}
  \Upsilon(\mu^2,{\rho'}_m)= \ & \frac{1}{2\pi i}\int_{\tilde{C}}
\frac{ds}{s}e^{{\rho'}_m/s}\frac{\mu
  s^{1/4}}{48}\left(((\mu s^{1/4})^3  \right. \\ 
& \ \ \ \ \ \ \ \ \ \ \ \ \ \ \ \ \ \left. +24 \mu
  s^{1/4})K_0(\mu s^{1/4})+(8(\mu s^{1/4})^2+48)K_1(\mu
  s^{1/4}) \right)
\end{split}
\nonumber
\end{equation}
and $F(s)=\tilde{F}(s^{1/k})$ with $k=4$.
Rescaling $s$ allows us to move all $\mu$ dependence into
${\rho'}_m(\mu)={\rho'}_m \mu^4$, so that we can write
$\Upsilon(\mu^2,{\rho'}_m)=\Upsilon({\rho'}_m(\mu))$. 
Unitarity ensures that $\hat{c}(\mu^2)\geq 0$ hence
\begin{equation}
  \label{bound1}
  \Delta c_{exact}-\Delta c_{est}\geq \min_{\mu} \Upsilon ({\rho'}_m(\mu))
\int_0^{\infty}d\mu^2
  \hat{c}(\mu^2),
\end{equation}
$\Delta c_{est}>0$ so that 
\begin{equation}
  \label{bound2}
  0 \geqslant \frac{\Delta c_{exact}-\Delta c_{est}}{\Delta c_{exact}}\geqslant
  \min_{\rho>0}\Upsilon (\rho).
\end{equation}
\begin{figure}
\epsfxsize=12cm
\epsfysize=5cm
\begin{center}
\epsffile{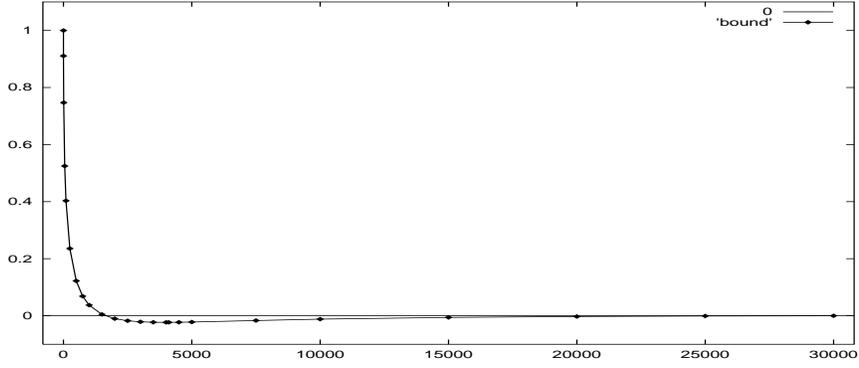}
\end{center}
\caption{Numerical integration of $\Upsilon (\rho)$ given in \eqref{psi}.}
\label{fig:bound}
\end{figure}
The lower bound in the relative undershoot of 
$I({\rho'}_m)$ therefore equals
$\min_{\rho}\Upsilon(\rho)$. 
In figure \ref{fig:bound} we plot $\Upsilon (\rho)$ for $k=4$. With
this value we get that $\min_\rho \Upsilon (\rho)=-0.0232$ 
so the relative overshoot in
$\Delta c_{est}$ compared with $\Delta c_{exact}$ is maximally $2.3\%$. It
follows from \eqref{bound1} that the bound in \eqref{bound2} is only
saturated for free theories (where $\hat{c}(\mu^2) \propto
\delta(\mu^2-m^2)$) for general interacting theories will the relative
overshoot be smaller than $|\min_\rho \Upsilon (\rho)|$. 
We show in figure \ref{fig:boundfig} the type of behaviour that we expect and which is
confirmed for the minimal models below.
\begin{figure}
\epsfxsize=11cm
\epsfysize=5cm
\begin{center}
\epsffile{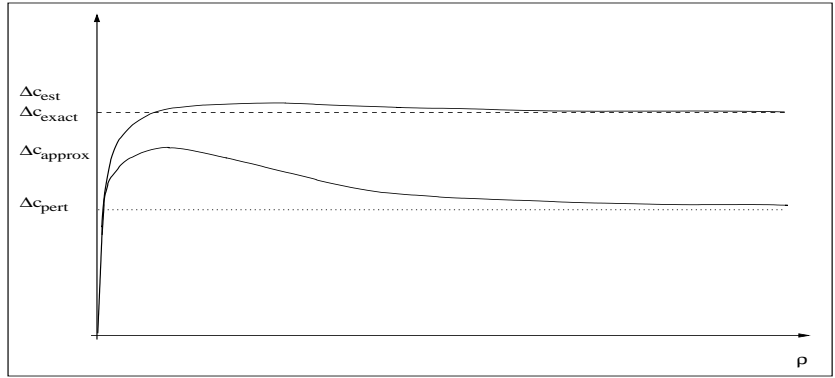}
\end{center}
\caption{The expected behaviour of $c_{UV}-I_n(\rho)$ and $c_{UV}-I(\rho)$.}
\label{fig:boundfig}
\end{figure}

The actual choice of $k=4$ then is a 
compromise between maximising the range of
analyticity (small $k$) and minimising the undershoot in $I({\rho'}_m)$ 
(large $k$).

\section{Application to different 2D models}
In this section we calculate the central charge for the free
bosonic and fermionic theory perturbed by a mass term.
These theories are gaussian and we can calculate $F(s)$ exactly, the
central charge can then 
be written as $\lim_{\rho \rightarrow \infty}I(\rho)$ as
the contribution from the cut then vanishes.  
The infra-red limit of these theories is trivial ($c_{IR}=0$) as all the
degrees of freedom are massive, and therefore decouple  
when approaching the infra-red fixed point where the scale goes to infinity. 
The free theories are none the less 
important to consider as we can here obtain the exact
function $I(\rho)$. 

In 4.3 we consider the unitary minimal models perturbed by the relevant
operator $\phi_{(1,3)}$, this theory has a non-trivial infra-red fixed point. 
We obtain the renormalisation group improved perturbative calculation of $c_{IR}$ and compare
with the approximation \eqref{approxcir}.

\subsection{The free boson}
We take the action for the free bosonic theory in 2 dimensions with a 
mass $m$ to be 
\begin{equation}
  \label{bosonvirk}
  S=\int d^2x\, \left(\frac{1}{2}\p_\mu\varphi(x)\p^\mu
  \varphi(x)+\frac{1}{2} m^2\varphi^2(x)\right) . 
\end{equation}
The perturbation away from the conformal field theory is thus given by the purely massive term
$\frac{1}{2}m^2\varphi^2$. The theory is still a free theory off criticality
and the correlator $\langle TT \rangle$ can be calculated exactly in the
whole scaling region from the ultra-violet to the infra-red.
With this normalisation the energy-momentum tensor becomes 
\begin{equation}
  \label{Tbose}
  T(z,\bar{z})=
T_{zz}(z,\bar{z})=-2\pi :\p\varphi (z,\bar{z})\p \varphi(z,\bar{z}):
\end{equation}
with the correlator 
\begin{eqnarray}
  \label{schwingerbose}
  \langle T(z,\bar{z})T(w,\bar{w})\rangle &=& (2\pi)^2\langle :\p
  \varphi(z,\bar{z})\p \varphi(z,\bar{z}):\, :\p
  \varphi(w,\bar{w})\p \varphi(w,\bar{w}): \rangle \nonumber \\ 
&=&  2 (2\pi)^2
\langle \p \varphi(z,\bar{z})\p \varphi(w,\bar{w})\rangle ^2,  
\end{eqnarray}
as only the double contractions survive. 
Using the form (\ref{boseG}) of the free propagator then (\ref{schwingerbose})
is $2(\p_z \p_z K_0(m|z|))^2$ where we have set $w=0$ using translation
invariance. We now use the identities $K_n'=-\frac{1}{2}(K_{n-1}+K_{n+1})$, 
$K_2(x)=K_0(x)+\frac{2}{x}K_1(x)$, $K_{-1}(x)=K_1(x)$ and  
(\ref{schwingerbose}) becomes
\begin{equation}
  \label{schwinger2}
 \langle T(z,\bar{z})T(0,0)\rangle =
  \frac{m^2 |z|^2}{8z^4}\left(
  4K_1^2(m|z|)\!+\! m^2 |z|^2 K_0^2(m|z|)\!+\! 4m|z| K_1(m|z|)K_0(m|z|)\right),
\nonumber
\end{equation}
hence $F(s)$ with $k=4$ becomes 
\begin{equation}
  \label{Fbose}
  F(s)=\left(\frac{ms^{1/2}}{2}\right)^2\left(\frac{4}{s^{1/2}}K_1^2(ms^{1/4}) +
       m^2K_0^2(m s^{1/4})+\frac{4m}{s^{1/4}}K_0(ms^{1/4})K_1(ms^{1/4})
       \right) .
\nonumber
\end{equation}
Knowing $F(s)$ exactly the central charges follow from
\eqref{Fgraense} directly, but let us use the relations
\eqref{align1}, \eqref{align2} in the limit of large $\rho$ where the cut
vanishes. For $c_{UV}$ we substitute $s'=ms^{1/4},\
\rho'=\frac{\rho}{m^4}$ in the contour $C_2$. 
From (\ref{align1}) we then get that (for $r_{IR}=\tfrac{1}{m^4}$)
\begin{equation}
  \label{cUVbose}
  I_{UV}(\rho)=\frac{1}{4\pi}\int_{-\frac{\pi}{2}}^{\frac{\pi}{2}} d\theta\, 
         e^{4i\theta}e^{\rho'e^{2i\theta}}\left(4e^{-2i\theta}
K_1^2(e^{i\theta})+K_0^2(e^{i\theta})+ 
4e^{-i\theta}K_0(e^{i\theta})K_1(e^{i\theta}) \right) .
\end{equation}
This is not expressible in terms of elementary functions, but it can be
calculated using the analytical properties of the Bessel functions. The
integration contour can be collapsed into a contour running along the
imaginary axis together with an infinitesimal semi-circle around the
origin. Setting $ix=e^{i\theta}$ we then get the 3 contributions
\begin{equation}
\frac{1}{4\pi i}\left(\int_{-1}^{-\epsilon}\!\!\!dx\!+\!\int_{\epsilon}^{1}\!\!\!dx\!+\!\int_{C_\epsilon}\!\!\!dx\right)
e^{-\rho x^2}\left(-4xK_1(ix)^2+x^3K_0(ix)^2-4ix^2 K_0(ix)K_1(ix) \right).
\nonumber
\end{equation}
For $\rho\rightarrow \infty$ only the contribution from the infinitesimal
semi-circle $C_\epsilon$ survive and here we can insert the asymptotic form for
$K_1(z)\sim z^{-1}$ and $K_0(z)\sim -\log z$. 
In this limit only the first term with $K^2_1$ will contribute as is seen 
from the asymptotic form.  
Taking into account that we only integrate over half a circle we then get the
well known result $c_{UV}=1$,
which is an exact result as the contribution from the cut vanishes in this
limit\footnote{We could also directly have used the analogous of
  \eqref{cirgranse} namely $\lim_{\rho \rightarrow
    \infty}I_{UV}(\rho)=\lim_{s\rightarrow 0}F(s)$ which is again seen
  changing variable in \eqref{align1} and taking the limit $\rho \rightarrow
  \infty$ in the integrand valid for all $r_{IR}<\infty$.}.
\begin{figure}
\epsfxsize=10cm
\epsfysize=9cm
\begin{center}
\epsffile{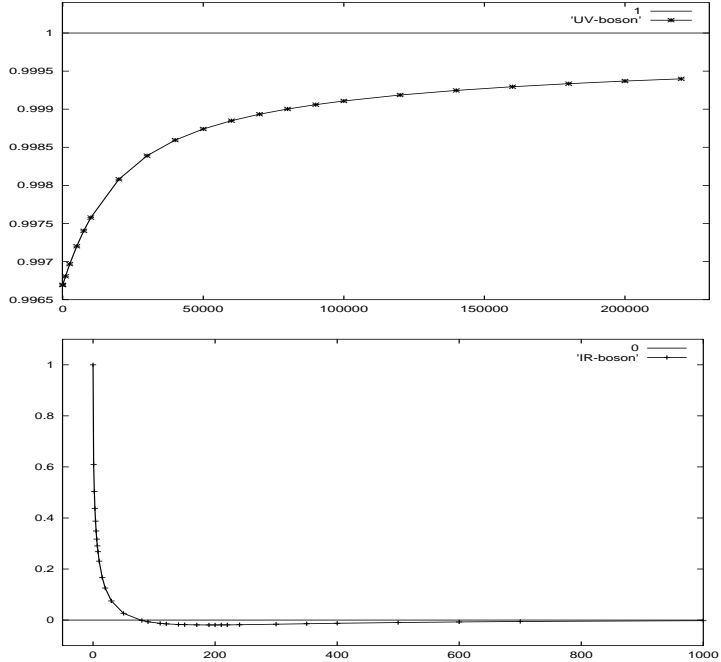}
\end{center}
\caption{a) Numerical integration of $I_{UV}(\rho)$ given by \eqref{cUVbose}
  compared with the exact value $c_{UV}=1$. b) Numerical integration of $I_{IR}(\rho)$.}
\label{fig:boseint}
\end{figure} 
To calculate the infra-red central charge we use \eqref{cirgranse}
hence $c_{IR}=\lim_{\rho
  \rightarrow \infty}I_{IR}(\rho)=\lim_{s\rightarrow \infty}F(s)$. From the
asymptotic form of the modified Bessel functions $K_n(x)\sim e^{-x}$,
together with \eqref{Fbose}, it follows that $c_{IR}=0$.
Figure \ref{fig:boseint}a 
shows the exact function $I_{UV}(\rho)$ computed by a   
numerical integration of 
\eqref{cUVbose} using a NAG Fortran Library integration routine. The figure
shows how the contribution from the cut vanishes in the limit of large $\rho$.
The exact infra-red function $I_{IR}(\rho)$ is plotted in figure \ref{fig:boseint}b,
it is important to note that the minimum value of $I_{IR}(\rho)$ is very
small, namely $-0.019$. 

\subsection{The free fermion}
The free massive fermion has the action
\begin{equation}
  \label{actionfermion}
  S=\int d^2x\, \left( \bar{\psi}\p
  \bar{\psi}+\psi\bar{\p}\psi+i m\bar{\psi}\psi \right) .
\end{equation}
A calculation analogous to the Bose case gives, for $k=4$,  that 
\begin{equation}
  \label{fermiF}
  F(s)=\frac{m^4s}{8}\left(K_1^2(ms^{1/4})\left(1+\frac{4}{m^2s^{1/2}}\right)-
  K_0^2(ms^{1/4})  \right) .
\end{equation}
The same arguments as above yields 
$c_{UV}=\tfrac{1}{2}$,  and the
infra-red contribution $c_{IR}=\lim_{s \rightarrow \infty}F(s)=0$, 
again by using the asymptotic form of the Bessel functions.
Again we can obtain the exact functions $I(\rho)$ by a numerical computation.
We get the same behaviour of $I_{IR}(\rho)$ 
as in the bosonic case, but with the 
minimum value $-0.0048$, so this exact function does again not differ
much from the infra-red central charge at its minimum value.

\subsection{The unitary minimal models}
The off critical quantum field theory picks out a specific renormalisation
 group flow from the ultra-violet to the infra-red
 conformal field theory.
If the quantum field theory is in the neighbourhood of one 
of the renormalisation group fixed points $\lambda^*$ 
in coupling  constant space (we may be choose the coordinates so 
$\lambda^*=0$ corresponds to the ultra-violet CFT) 
the action can be written as  
\begin{equation}
S=S_{CFT}+\sum_{i=1}^N\lambda^i \int d^2 x\ \Phi_i(x). 
\label{generelvirkning}
\end{equation}
Here the $\Phi_i(x)$'s are scaling fields with scaling dimension $\Delta_i$,
the coupling constants $\lambda^i$ then 
have mass dimension
$[\lambda^i]=2-\Delta_i=y_i$.  From (\ref{generelvirkning}) it 
follows that $y_i$ is the renormalisation group eigenvalue of the scaling fields $\Phi_i$. For
renormalisable quantum field theories we need $y_i\geq 0$ and the scaling operators therefore
have to be relevant $(y>0)$ or marginal $(y=0)$. For relevant operators we
will move away from the fixed point when the scale increases and $S_{CFT}$ thus
corresponds to an ultra-violet critical point. 

We consider quantum field theories which have the unitary minimal models ${\cal M}(m)$ 
  as their scaling limits.  
The central charge for ${\cal M}(m)$ is given by 
\begin{equation}
c(m)=1-\frac{6}{m(m+1)},\ \ \ m=3,4,\ldots 
\label{minmodc}
\end{equation}
There are  $m(m-1)/2$ primary fields 
$$
\phi_{(p,q)} =\phi_{(m-p,m-q+1)},\ \ \ 1\leq p\leq m-1,\ \  1\leq q \leq m,
$$ 
with the conformal dimensions
\begin{equation}
h_{(p,q)}=\bar{h}_{(p,q)}=\frac{((m+1)p-mq)^2-1}{4m(m+1)},
\label{conformaldmin}
\end{equation}  
so the primary fields are scalars (spin zero). 
It follows from (\ref{conformaldmin}) that
$2(m-2)$ of the primary fields 
satisfy $\Delta=2h<2$ and are thus relevant operators 
\footnote{As the secondary fields will have a conformal dimension of at least
  $h+1$ there will be no relevant secondary fields.}.
We will consider the minimal models perturbed by the relevant operator
$\phi_{(1,3)}$  given by the action
\begin{equation}
S=S_{{\cal M}(m)}-\lambda_0 \int d^2x\ \phi_{(1,3)}(x).
\label{13virkning}
\end{equation}
The reasons for choosing this operator are that:\footnote{The model
  \eqref{13virkning} is integrable and was first studied in
  \cite{intezam}. The only other integrable perturbations of the unitary
  minimal models are with $\phi_{(1,2)}$ and $\phi_{(2,1)}$. These three
  models corresponds respectively to the Korteweg-de Vries,
  Gibbon-Samede-Kotera and Kupersmidth equations \cite{watts}.}\\ 
{\it i})  $\phi_{(1,3)}$ is a relevant field, $h_{(1,3)}=1-\frac{2}{m+1}<1$,
which exists in all ${\cal  M}(m)$.\\  
{\it ii}) ${\phi_{(m,n)}}$ form an algebra under the operator product
expansion, from the fusion rules of minimal models it follows that the set
${\phi_{(1,n)}}$ constitutes a sub-algebra in which only $\phi_{(1,1)}$, 
$\phi_{(1,2)}$ and $\phi_{(1,3)}$ are relevant as seen from
\eqref{conformaldmin}. $\phi_{(1,3)}$ is normalised so that the structure
constant $C_{(1,3)(1,3)(1,1)}=1$ where $I=\phi_{(1,1)}=\phi_{(m-1,m)}$ is the
identity, and it has the self coupling $C_{(1,3)(1,3)(1,3)}=b(m)$. It does
not couple to $\phi_{(1,2)}$ so it has no coupling to other relevant
operators in the sub-algebra ${\phi_{(1,n)}}$ \cite{zam2,cardy2,dotsenko}. This
means that there is a renormalisation group flow connecting the ultra-violet and infra-red fixed points along the
direction of $\phi_{(1,3)}$ so it is consistent to include only the one
relevant field $\phi_{(1,3)}$\footnote{This can be seen by writing the Zamolodchikov metric in normal
  coordinates around the ultra-violet fixed point $g=0$, $G_{ij}=\delta_{ij}+O(g^2)$, the
  beta-functions become in these coordinates $\beta^i(g)=-y^ig_i-\pi 
\sum_{j,k}C^i_{jk}g^jg^k+O(g^3)$. If $g^i\neq 0$ only for $g^{(1,3)}$ and
$C_{(1,3)(1,3)j}=0$ for $j\neq (1,3)$ then $\beta^{j}(g^{(1,3)})=0$ for
$j\neq (1,3)$ and there is no flow transverse to the $\phi_{(1,3)}$ direction
\cite{cappellipert}.}, i.e.\ it is a geodesic renormalisation group trajectory \cite{laessig}.\\
{\it iii}) $\phi_{(1,3)}$ is the least relevant field, and the perturbation
in  (\ref{13virkning}) becomes marginal in the
limit of $m\rightarrow \infty$ as $y=\frac{4}{m+1}$ so that 
$y\rightarrow 0$.  In this limit the
fixed points are arbitrarily close in coupling constant space, 
and perturbation theory is viable in the whole region from
the ultra-violet to the infra-red.  

We want to calculate the difference between the ultra-violet and the
infra-red central charge $\Delta c=c_{UV}-c_{IR}$.
It has been argued that the infra-red conformal field theory of
\eqref{13virkning} is given by the unitary minimal model 
${\cal M}(m-1)$, as $\Delta c$ in the perturbative limit $y\rightarrow 0$ is
given by $\Delta c=\tfrac{3}{16}y^3+O(y^4)$ \cite{zam2,cardy2}, 
and from (\ref{minmodc}) we get that 
\begin{equation}
c(m)-c(m-1)=\frac{12}{m(m^2-1)}=\frac{3y^3}{2(2-y)(4-y)}
=\frac{3 y^3}{16}+O(y^4).
\label{deltacmin}
\end{equation}  
A general argument for all $m$ has been given by a thermodynamic Bethe ansatz
method in \cite{alzam}\footnote{In \cite{morris} the first fixed points 
${\cal M}(m)$, $m=3,4,...,12$ were found numerically using the
exact renormalisation group.}. We now describe how $\Delta c$ is calculated
using our approximation method\footnote{Another way of calculating $\Delta c$
follows from the proof of the $c$-theorem \cite{zamolodchikov}, where it is
shown that $R^2\tfrac{dC}{dR^2}=-\tfrac{3}{4}R^2\langle T^\mu_\mu
T^\mu_\mu\rangle$ leading to Cardy's sum rule \cite{cardysum} $\Delta
c=-\tfrac{3}{4}\int_0^\infty d(R^2)R^2 \langle T^\mu_\mu T^\mu_\mu
\rangle$. The correlator is normally only known perturbatively so this gives
a perturbative estimate of $\Delta c$.}, 
and we will compare this with the exact
result $\Delta c_{exact}=c(m)-c(m-1)$.

\begin{figure}
\epsfxsize=12cm
\epsfysize=6cm
\begin{center}
\epsffile{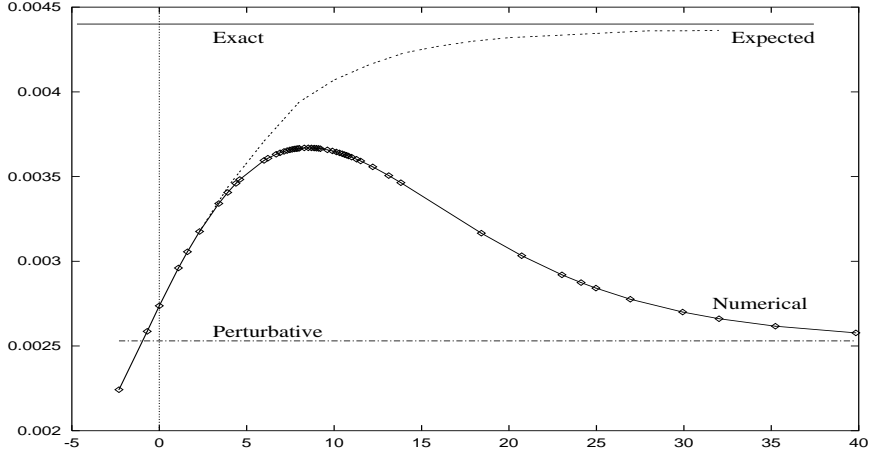}
\end{center}
\caption{The numerical-result $c_{UV}-\tilde I_1(\rho')$ against
  $\log{\rho'}$ for $m=14$. Also plotted is the exact and perturbative values
  $\Delta c_{exact}$ and $\Delta c_{pert}$. 
The dashed line is the expected behaviour of $\tilde{I}(0)-\tilde{I}(\rho')$.}
\label{fig:resulm14}
\end{figure}

\begin{figure}
\epsfxsize=12cm
\epsfysize=6cm
\begin{center}
\epsffile{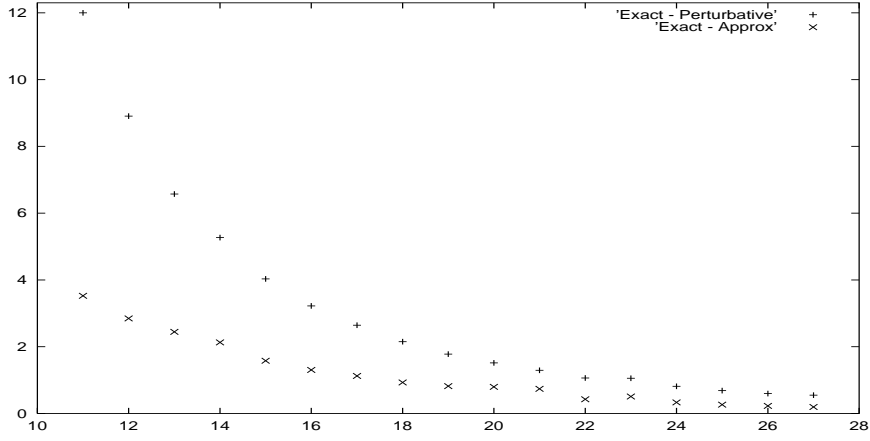}
\end{center}
\vspace{-0.2cm}
\caption{$(\Delta c_{exact}-\Delta c_{approx})m(m^2-1)$ and $(\Delta
  c_{exact}-
\Delta c_{pert})m(m^2-1)$ 
against  $m$.}
\label{fig:error}
\end{figure}

To construct the term $\langle T(z,\bar{z})T(0,0)\rangle$ we use
the Ward identities which follow from euclidean invariance
\begin{eqnarray}
& & \p_{\bar{z}}\langle T(z,\bar{z})\Phi_1(x_1)\cdots
\Phi_n(x_n)\rangle+\frac{1}{4}\p_z \langle \Theta(z,\bar{z})\Phi_1(x_1)\cdots
\Phi_n(x_n) \rangle \nonumber \\ 
&=& \pi \sum_{i=1}^n
(\delta(z-x_i)\p_{x_i}-\p_z\delta(z-x_i)h_i) \langle \Phi_1(x_1)\cdots
\Phi_n(x_n) \rangle,
\label{wardidentities}
\end{eqnarray}
here $\Phi_i$ are primary fields with conformal dimension $h_i$. For the
correlator we are interested in contact terms vanish, and we get 
\begin{equation}
\p_{\bar{z}_1}\p_{\bar{z}_2}\langle T(z_1,\bar{z}_1)T(z_2,\bar{z}_2)\rangle =
\frac{1}{4^2}\p_{z_1}\p_{z_2}\langle
\Theta(z_1,\bar{z}_1)\Theta(z_2,\bar{z}_2)\rangle .
\label{diffeqgeneral}
\end{equation} 
$\Theta$ is the infinitesimal generator for scale transformation (hence its
vanishing in the CFT), 
and in a renormalisable field theory $\Theta$ must belong to the
space spanned by the relevant and marginal fields defining the
perturbation away from criticality in (\ref{generelvirkning})
\begin{equation}
\Theta(x)\equiv 2\pi \sum_{i=1}^N\beta^i(g)\Phi_i(x),
\label{defbeta}
\end{equation}
where $\beta^i(g)$ is the beta-function given in terms of the renormalised
coupling constants $g$ \cite{zam2}. 
(\ref{diffeqgeneral}) can thus be written
\begin{equation}
\p_{\bar{z}_1}\p_{\bar{z}_2}\langle
T(z_1,\bar{z}_1)T(z_2,\bar{z}_2)\rangle
=\frac{\pi^2}{4}\beta^i(g)\beta^j(g)\p_{z_1}\p_{z_2}\langle \Phi_i (z_1,\bar{z}_1) 
\Phi_j (z_2,\bar{z}_2) \rangle.
\label{diffeq1}
\end{equation}
The correlator $\langle \Phi_i \Phi_j \rangle$ can be calculated in
perturbative conformal field theory using the operator product expansion in
the ultra-violet conformal field theory.
The bare correlator is in the lowest order in $\lambda_0$ given by
\cite{cappellipert}\footnote{conn.\ stands for the connected correlator.}
\begin{eqnarray}
\langle \phi(x) \phi(0) \rangle &=&
\frac{\langle\phi(x)\phi(0)e^{\lambda_0\int d^2x' \phi(x')} \rangle_{{\cal M}(m)}}{\langle
  e^{\lambda_0\int d^2x'  \phi(x')}\rangle_{{\cal M}(m)}}\nonumber \\
& = &  \langle \phi(x)\phi(0)\rangle_{{\cal M}(m)}+\lambda_0\int d^2x' \langle
\phi(x)\phi(0)\phi(x')\rangle_{{\cal M}(m),\mbox{\small{conn.}}}
 +O(\lambda_0^2) \nonumber \\ 
&=&\frac{1}{| x |^{2(2-y)}}\left(1+\lambda_0\frac{4\pi b(y)
     A(y)}{y}| x |^y+O(\lambda_0^2)\right),
\end{eqnarray}
where $\phi(x)$ is the bare field $\phi_{(1,3)}(x)$ and 
$A(y)=\frac{\Gamma (1-y)\Gamma (1+y/2)^2}{\Gamma (1-y/2)^2 \Gamma (1+y)} 
=1+O(y^3)$. The operator product expansion coefficient $b(y)$ can be
calculated from a Coulomb gas representation of the minimal models using the
formulas in \cite{dotsenko}
\begin{equation}
b(y)^2=\tfrac{16}{3}\tfrac{(1-y)^4}{(1-y/2)^2(1-3y/4)^2}
\left(\tfrac{\Gamma(1+y/2)}{\Gamma(1-y/2)}\right)^4\left(
\tfrac{\Gamma(1-y/4)}{\Gamma(1+y/4)}\right)^3\left(\tfrac{\Gamma(1-y)}{\Gamma(1+y)}\right)^2\left(\tfrac{\Gamma(1+3y/4)}{\Gamma(1-3y/4)}\right)=\tfrac{16}{3}+O(y).
\nonumber
\end{equation}
Choosing the renormalisation conditions $\langle
\phi(x,g)\phi(0,g)\rangle|_{\mid x \mid=\mu^{-1}}\equiv \mu^4$, the
renormalised correlator and the $\beta$-function becomes \cite{cappellipert}
\begin{eqnarray}
\langle \phi(x,g)\phi(0,g)\rangle &=&\frac{\mu^4}{|\mu x
  |^{2(2-y)}}\left(1+\frac{4\pi A(y) b(y) g}{y}(| \mu x
  |^y-1)+O(g^2)\right),\nonumber \\  
 \beta(g) &=& -yg-\pi b(y) g^2 A(y)+O(g^3),
\label{phiphi}
\end{eqnarray}
where $\phi(x,g)$ is the renormalised field and $g$ is the renormalised
coupling. 
The zeros of the $\beta-$function, the renormalisation group fixed points,  
are thus $g_{UV}=0$, $g_{IR}^*=\tfrac{-y}{\pi A(y)b(y)}$ and
therefore $g\in (\tfrac{-y}{\pi A(y)b(y)},0 )$ as the theory
\eqref{generelvirkning} lies between the two scaling limits.
From the Callan--Symanzik equation we get the running coupling constant \cite{cappellipert}
\begin{equation}
  \label{running}
\bar{g}(|x|)=|\mu x|^y\frac{g}{1-\frac{\pi A(y)b(y) g}{y}(|x\mu |^y-1)},  
\end{equation}
interpolating between $g_{UV}$ for $|x|\rightarrow 0$ and $g_{IR}^*$ for
$|x|\rightarrow \infty$ and satisfying $\bar{g}(\mu^{-1})=g$.
Euclidean invariance 
allows us to write the correlator of the
energy-momentum tensor as $\langle
T(z,\bar{z})T(0,0)\rangle =\frac{\tilde{F}(\tilde{R})}{2 z^4}$ in terms of the
dimensionless quantity $\tilde{R}=\mu^2 z \bar{z}$. The differential equation
\eqref{diffeq1} then becomes
\begin{equation}
  \label{diffeq5}
  \frac{\p ^2}{\p \tilde{R}^2}\tilde{F}(\tilde{R})=\frac{\pi^2 \beta^2}{4\mu^4}
  \tilde{R}^2 \frac{\p^2}{\p \tilde{R}^2}\langle \phi(\tilde{R}) \phi(0) \rangle .
\end{equation}
A solution to this equation is given by\footnote{These equations directly
  generalises to the case with more couplings.}
\begin{equation}
  \label{partsol}
  \tilde{F}(\tilde{R})=\frac{\pi^2 \beta^2}{2\mu^4}
\left(\tilde{R}^2\, \langle \phi \phi\rangle-4\tilde{R}
  \int^{\tilde{R}}\! d\tilde{R}'\, \langle \phi \phi
  \rangle+6\int^{\tilde{R}}\! d\tilde{R}'\int^{\tilde{R}'}\! d\tilde{R}''\, \langle
  \phi \phi \rangle  \right)+\alpha_1+\alpha_2 \tilde{R},
\nonumber
\end{equation}
where $\alpha_1, \alpha_2 \in \mathbb{R}$. The
differential equation \eqref{diffeq5} is a boundary value problem as
$\tilde{F}(\tilde{R})$ is known in the scaling limits
\begin{equation}
\tilde{F}(\tilde{R})\rightarrow \left\{ 
\begin{array}{ll}
c_{UV} & \mbox{for }\tilde{R}\rightarrow 0. \\
c_{IR}^* & \mbox{for }\tilde{R}\rightarrow \infty. 
\end{array} \right.
\end{equation}
In the limit where $\tilde{R}\rightarrow 0$ the correlator $\langle \phi \phi
\rangle$ scales as in ${\cal M}(m)$ i.e.\ $\langle \phi \phi\rangle \sim 
\tfrac{1}{\tilde{R}^{2h}}$, hence $\tilde{R}^2 \langle \phi
\phi\rangle \sim \tilde{R}^{y}\rightarrow 0$,
$\tilde{R}\int^{\tilde{R}}d\tilde{R}'\langle \phi \phi \rangle \sim
\tilde{R}^y \rightarrow 0$, and finally
$\int^{\tilde{R}}d\tilde{R}'\int^{\tilde{R}'}d\tilde{R}'' \langle \phi \phi
\rangle \sim \tilde{R}^y\rightarrow 0$ for $\tilde{R}\rightarrow 0$. This sets
the boundary value $\alpha_1=c_{UV}$. As $\tilde{F}(\tilde{R})$
attains a
finite value for $\tilde{R}\rightarrow \infty$ then all linear terms in
$F(\tilde{R})$ must cancel and we set $\alpha_2=0$.
Integrating \eqref{phiphi} and inserting the boundary conditions gives
\begin{equation}\label{intphiphi}
\begin{split}
  \tilde{F}(\tilde{R})=&
c_{UV}+\frac{\pi^2 g^2
  \tilde{R}^y}{2}\left(\frac{y(2-y)(3-y)}{y-1}+2\pi
  A(y)b(y)g\left( \frac{(2-y)(3-y)}{1-y} \right. \right. \\ 
& \left. \left. +\tilde{R}^{\frac{y}{2}}\frac{(3y-4)(3y-6)}{3(\frac{3}{2}y-1)}\right) \right).
\end{split}
\end{equation}
Determining the function $F(s)$ as $F(s)=
\tilde{F}(\tilde{R})|_{\tilde{R}=1,g=\bar{g}(s)}$,
the theory is fixed at the point of renormalisation
$|x|^{-1}=\mu$, and all the scale dependence is moved into the running coupling
constant. The running coupling becomes  with $|x|\rightarrow s^{1/4y}$:    
\begin{equation}
  \label{runnings}
  \bar{g}(s)=\frac{g s^{\frac{1}{4}}}{1-\frac{\pi
  A(y)b(y)g}{y}(s^{\frac{1}{4}}-1)}, 
\end{equation}
so that scale-transformations  move around in the coupling constant space.  
The 1 loop renormalisation group improved approximation to $F(s)$ then becomes
\begin{equation}  \label{approxf}
\begin{split} 
 F_1(s)=&\ c_{UV}+\frac{\pi^2}{2}\bar{g}^2(s)\left(
\frac{y(2-y)(3-y)}{y-1} \right. \\ 
&\left.\ +2\pi A(y)b(y)\bar{g}(s)\left(
\frac{(2-y)(3-y)}{1-y}+\frac{(3y-4)(3y-6)}{3(\frac{3}{2}y-1)} \right) \right).
\end{split}
\end{equation}
Here $k=4$ as $F_1(s)=\tilde{F}_1(s^{1/4})$.
To obtain the approximation for $c_{IR}$ we then have to calculate
the contour integral
\begin{equation}
I_1(\rho)=\frac{1}{2\pi i}\int_{C_0}ds\frac{e^{\rho /s}}{s}F_1(s).
\label{52}
\end{equation}
In the limit $m\rightarrow \infty$ the ultra-violet and infra-red fixed points are
perturbatively close in coupling constant space as noted above, 
hence $F_1(s)$ correctly describes $F(s)$ in this limit 
and we should take $\rho\rightarrow
\infty$ in \eqref{align2} thus eliminating the contribution from the
cut. 
The approximation then becomes $\lim_{\rho \rightarrow \infty}I_1(\rho)=\lim_{s \rightarrow
  \infty}F_1(s)=c_{IR}^*$ using \eqref{graense1}, and the approximation 
in this limit thus equals the RG improved
perturbative result which is 
\begin{equation}\label{delcpert}
 \begin{split}
 \Delta c_{pert} 
  =&\,
 c_{UV}-c_{IR}^* =
 - \frac{\pi^2}{2}(g_{IR}^*)^2\left(
 \frac{y(2-y)(3-y)}{y-1} \right. \\ 
 &\left.\ +2\pi A(y)b(y)g_{IR}^*\left(
 \frac{(2-y)(3-y)}{1-y}+\frac{(3y-4)(3y-6)}{3(\frac{3}{2}y-1)} \right) \right)
  \\ 
 =&\, \frac{3y^3}{16}+O(y^4).
 \end{split}
 \end{equation}
This is equal to the asymptotic form of the exact value $c(m)-c(m-1)$ in
\eqref{deltacmin}. We wish to improve this result using 
the approximation \eqref{approxcir}.
We rewrite the running coupling constant
\begin{equation}
  \label{runnings2}
  \bar{g}(s)=\frac{g s^{1/4}}{1-\frac{\pi
  A(y)b(y)g}{y}(s^{1/4}-1)}=\frac{g_{IR}^*
  |\tilde{g}|s^{1/4}}{1+|\tilde{g}|s^{1/4}},\ \ \ \ \ \ 
\tilde{g}=\frac{g}{g-g_{IR}^*}\ \in \ \ ( -\infty,0). 
\end{equation}
All dependence of the renormalised coupling $g$ are now moved into the
parameter $\rho$ setting $s'=s|\tilde{g}|^{4}$ and 
$\rho'=\rho |\tilde{g}|^{4}$
\begin{equation}
  \label{runningc2}
  I_1(\rho,g)=\frac{1}{2\pi i}\int_{C_0}ds \frac{e^{\rho/ s}}{s}F_1(s^{})=
\frac{1}{2\pi i}\int_{C'_0}ds' 
\frac{e^{\rho'/s'}}{s'}F_1({s'})=\tilde{I}_1(\rho')
\end{equation}
and then $\bar{g}(s')=\tfrac{g_{IR}^*}{1+{s'}^{1/4}}$. 
This contour integral can be evaluated for example doing a numerical
integration or a series expansion in $F_1(s')$, we have done both. 
A numerical integration of \eqref{runningc2} with $m=14$ using the NAG Fortran
Library is shown in figure \ref{fig:resulm14}.
In figure \ref{fig:resulm14} the dashed line indicates the expected behaviour of the exact
function $\tilde{I}(0)-\tilde{I}(\rho')$ which is taken analogous to 
the curves of the free theories in figure \ref{fig:boseint}. 
Writing $F_1(s')$ as a power series 
\begin{equation}
F_1(s')=\Phi_1(\bar{g}(s'))=\sum_{n=0}^{\infty} h_n (s')^{ n/4}
\label{power}
\end{equation}
and inserting this into \eqref{runningc2} and using the integral
representation 
\begin{equation}
  \label{gamma}
\frac{1}{\Gamma(1+z)}=\frac{1}{2\pi i}\int_{{C'}_0}ds'
\frac{e^{1/s'}}{s'} (s')^z , 
\end{equation}
we then obtain 
\begin{equation}
  \label{gammaapprox}
  \tilde{I}_1(\rho')=\sum_{n=0}^{\infty} \frac{h_n (\rho')^{n/4}}
   {\Gamma (1+n/4)}
\end{equation}
which is recognised as the Borel transform of order $k=4$.
This expression can be computed numerically (using e.g.\ Maple) 
by truncating to a finite $n$, and the minimal value can be found.
\begin{center}
\begin{tabular}{| c || c | c | c | c |}
\hline 
m & $\Delta c_{exact}$ & $\Delta c_{pert}$ & $\Delta c_{approx}$ & $\Delta
c_{approx2}$ \\
\hline \hline 
11    & 0.00909     & 0        &   0.00642  & 0.00970  
\\ \hline       
12 & 0.00699     & 0.00180    & 0.00533     & 0.00721
\\ \hline
13 & 0.00549     &  0.00248   & 0.00437     & 0.00556
\\ \hline
14 & 0.00440     &  0.00253   & 0.00368     & 0.00440
\\ \hline
15 & 0.00357     &  0.00237   & 0.00310     & 0.00357 
\\ \hline
16 & 0.00294     &  0.00215   & 0.00262     & 0.00293
\\ \hline
17 & 0.00245     &  0.00191   & 0.00222     & 0.00244
\\ \hline
18 & 0.00206     &  0.00169   & 0.00190     & 0.00205
\\ \hline
19 & 0.00175     &  0.00149   & 0.00163     & 0.00175
\\ \hline
20 & 0.00150     &  0.00131   & 0.00140     & 0.00150
\\ \hline
21 & 0.00130     &  0.00116   & 0.00122     & 0.00130
\\ \hline
22 & 0.00113     &  0.00103   & 0.00109     & 0.00113
\\ \hline
23 & 0.000988    &  0.000911   & 0.000956     & 0.000989
\\ \hline
24 & 0.000870    &  0.000811   & 0.000846     & 0.000871
\\ \hline
25 & 0.000769    &  0.000725   & 0.000752     & 0.000771
\\ \hline  
26 & 0.000684    & 0.000650    & 0.000671     & 0.000686
\\ \hline 
27 & 0.000611  & 0.000583       & 0.000601    & 0.000613
\\ \hline
\end{tabular} 
\end{center}
In the table the obtained values of the numerical
integration denoted $\Delta c_{approx}$ are listed 
together with the exact results $\Delta c_{exact}$ and the RG improved perturbative
values $\Delta c_{pert}$. For $m<11$ the perturbative result will 
break down as $\Delta c_{pert}$ becomes negative and thereby violates unitarity.
In \cite{lkpm} we used $k=2/y=(m+1)/2$, 
but here we have chosen $k=4$ 
as we then get a stricter bound on the exact function as explained in section
3.3. The results for $k=2/y$ and $k=4$ are similar for all $m$
calculated,  except $m=11, 12$ where they differ slightly. 
The improvement of the approximation \eqref{approxcir} over the RG
improved 
perturbative result is seen to be significant. In figure \ref{fig:error} $\Delta
c_{exact}-\Delta c_{approx}$ and $\Delta c_{exact}-\Delta c_{pert}$ are  
plotted (scaled with $m(m^2-1)$ so that all the
points can be distinguished) against $m$. The horizontal axis is then the
exact value. The figure shows that the
approximation improves the perturbative results 
with more than a factor two.

\begin{figure}
\epsfxsize=12cm
\epsfysize=6cm
\begin{center}
\epsffile{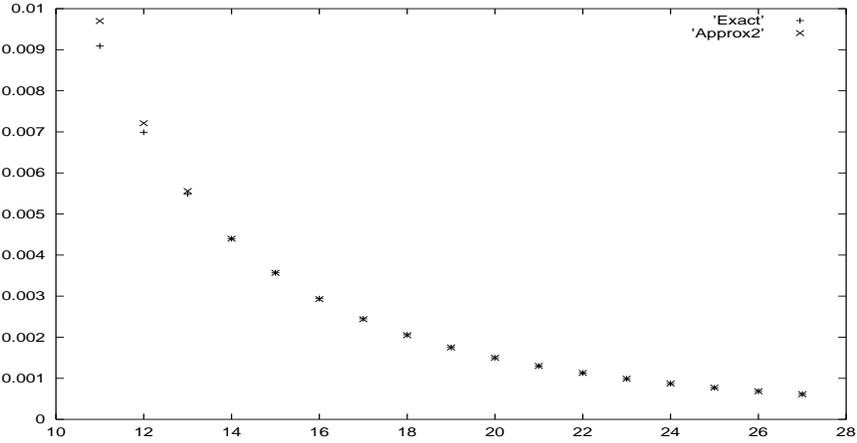}
\end{center}
\caption{$\Delta c_{exact}$ and $\Delta c_{approx2}$ against $m$.}
\label{fig:approx2}
\end{figure}

If we choose the maximum value $k=1$ we get 
results which are almost
identical to the exact values, we denote these by $\Delta c_{approx2}$, and
plot in figure \ref{fig:approx2} $\Delta c_{exact}$ and $\Delta c_{approx2}$ against $m$,
the numbers are also listed in the table. In this case though we do not have
 a strict bound on the exact function
$I(\rho)$ as for the case with $k=4$ described in section 3.3. The exact
function might still have a very small undershoot if for example the smallest
non zero mass of a one particle state is small compared with $\rho'_m$, 
i.e.\ $\Delta c_{est}\sim \Delta
c_{exact}$. Figure \ref{fig:approx2} shows the very 
good correspondence between
$\Delta
c_{exact}$ and $\Delta c_{approx2}$, this is remarkable because 
the approximation is based on only a one loop calculation. 

\section{Critical exponents for $\varphi^4$ in 3 dimensions}
The other example we will consider is $\varphi^4$ theory in 3 dimensions
with $O(N)$ symmetry\footnote{This model is one of the most studied models in
critical phenomena and it is important because it shares its infra-red fixed point
with a number of physical models, such as: polymers ($N=0$), the Ising
model ($N=1$), super-fluid Bose-liquid ($N=2$) and the Heisenberg ferromagnet
($N=3$).}.  We will study the Ising ($N=1$) case in detail here,
but the method applies for a general $N$.
Quantum field theories have generic infra-red divergences in dimensions lower than 4, for 
$\varphi^4$ theory this can be seen by studying the 1PI 2 point function
\cite{parisibook}. A way of regulating these infra-red divergences is to either do
an $\epsilon$ expansion in $\epsilon=4-D$, or work with a massive theory. We
will here do the latter, and work in $D=3$ following \cite{parisipaper}. We
introduce the renormalised field, mass and coupling according to the
conventions in \cite{IogD} so the renormalised fields $\varphi^R$ 
are given as: $\varphi
=(Z_1)^{1/2} \varphi ^R$, $\varphi^2= Z_2 (\varphi^2)^R$ and $m, g$ denotes the
renormalised mass and coupling. 
The infra-red divergences are now removed and only show up in the bare correlators
as non analytic dependence in the bare coupling \cite{parisipaper}.  
We will calculate two of the infra-red critical exponents $\nu$ and $\eta$, all other
exponents follow from scaling relations\footnote{The calculation of the critical exponents in $\varphi^4$ theory 
are among the most precise calculations in quantum field theory,
and they have been calculated using both the $\epsilon$ expansion, exact 
renormalisation group
arguments, perturbative quantum field theory, high temperature expansions,
strong coupling expansions 
and monte Carlo simulations. 
These exponents describe the scaling behaviour of the
theory in the scaling region and are universal in the
sense that they are only determined by the infra-red fixed point and are
therefore identical for all quantum field theories flowing to this point under the renormalisation group,
irrespective of the underlying microscopic dynamics.}, 
e.g.\ Fisher's scaling relation $\gamma=\nu (2-\eta)$.

The free 2 point correlator, or propagator, is given as
\begin{equation}
  \label{propagator1}
  G_{0}(p,m)=\int d^D x\ e^{i p x}\langle \varphi(x)\varphi(0) \rangle=\frac{1}{p^2+m^2}.
\end{equation}
The mass is related to the critical temperature $\theta$ as $\theta\sim m^2$,
so that the free propagator can be written as
\begin{equation}
  \label{propagator2}
  G_0(p,\theta)=\theta^{-\gamma}h(\tfrac{p}{\theta^\nu})
\end{equation}
where $h$ is regular at the origin and  
$\gamma=1$ and $\nu=\tfrac{1}{2}$. 
This scaling behaviour generalises to the interacting theory with critical
exponents $\nu$ and $\gamma$, hence from the scaling relations all
critical exponents can be obtained from the propagator.
We will now define the function $\tilde{F}(s)$ above for respectively $\nu$ and
$\eta$, and show that they are analytic in some sector with the
correct scaling limits.

Using that the massless theory for small momentum (infra-red region) is equivalent
to the large momentum region at $g=g_{c}$ \cite{parisipaper}
it follows from the Callan-Symanzik equation for the renormalised
propagator that in the limit of small momentum it scales as 
$ G(p)\sim \frac{1}{p^{2-\eta}}$, hence defining 
\begin{equation}
  \label{nu1}
  \tilde{F}_{\eta}(p)=p\frac{\p G_2^R(p)}{\p p}/G^R_2(p)+2, 
\end{equation}
would satisfy the infra-red behaviour $\lim_{p\rightarrow
  0}\tilde{F}_{\eta}(p)=\eta$. To get the ultra-violet behaviour we can look at the
spectral representation
\begin{equation}
  \label{spectra1}
\tilde{F}_{\eta}(p)=-2\ \frac{\int_0^\infty \!d\mu^2\
  \tilde{c}(\mu^2,m,g)\frac{p^2}{(p^2+\mu^2)^2}}{\int_0^\infty\! d\mu^2\ \tilde{c}(\mu^2,m,g)
\frac{1}{p^2+\mu^2}}+2\rightarrow 0 \ \ \mbox{  for
  }\  p\rightarrow \infty,
\end{equation}
or simply use that $G_2^R(p)\rightarrow \tfrac{1}{p^2+m^2}$ in the
ultra-violet (the spectral density will only contribute with $\delta (\mu^2-m^2)$),
showing that $\lim_{s\rightarrow \infty}\tilde{F}_{\eta}(s)=0$, as it
should be using the scaling relation $\nu(2-\eta)=\gamma$.
From this spectral decomposition it also directly follows that $\tilde{F}_{\eta}(s)$ is
analytic for $s\in S(\pi-\epsilon')$ for $\epsilon'\ll 1$ \footnote{Actually it will be analytic in $S(\pi)$ but
  to get the Borel transform $\tilde{F}_{\eta}(s)$ needs to be bounded at the origin so that
  $\tilde{F}_{\eta}(s)$ have to be bounded for all closed subsets of $S$ which would not be
  the case with $S(\pi)$.}, and that 
\begin{equation}
  \label{limitsF1}
\tilde{F}_{\eta}(s)\rightarrow \left\{ 
\begin{array}{ll}
\eta & \mbox{for }s\rightarrow 0_{+}, \\
0 & \mbox{for }s\rightarrow \infty. 
\end{array} \right.  
\end{equation}
Dimensional analysis gives us that 
\begin{equation}
  \label{dimanal}
  -m\frac{\p G_2^R(p,m)}{\p m}=p\frac{\p G_2^R(p,m)}{\p p}-[G_2]G_2^R(p,m)
=\left(p\frac{\p}{\p p}+2\right)G_2^R(p,m)
\nonumber
\end{equation}
hence it follows that 
\begin{equation}
\tilde{F}_{\eta}(p)=-m\frac{\p G_2^R(p,m)}{\p m}/G_2^R(p,m)
=Z_1^{-1}m\frac{\p Z_1}{\p  m},
\end{equation}
also showing that the ultra-violet value is zero as the gaussian theory is ultra-violet
finite. 
We also want to compute $\nu$, it again follows from the Callan--Symanzik equation for the $n$ point Green
function with $s$ insertions of $\varphi^2$ that the scaling behaviour for
low momenta at the critical point is \cite{IogD}
\begin{equation}
  \label{21scaling}
  G_{n,s}^R(\lambda p,\lambda q,m)\sim \lambda^{D-\tfrac{1}{2}n(D+2-\eta)-\frac{s}{\nu}}.
\end{equation}
Hence in the limit $p\rightarrow 0$
\begin{equation}
  \label{gamm2def}
  p\frac{\p G_{2,1}^R(p,q,m)}{\p p}/G_{2,1}^R(p,q,m)=-2+\eta-\frac{1}{\nu},
\end{equation}
so we define 
\begin{equation}
  \label{F2gamma2}
  \tilde{F}_{\nu}(p)=-\tilde{F}_{\eta}(p)+4+p\frac{\p G_{2,1}^R(p,-p,q=0,m)}{\p p}/G_{2,1}^R(p,-p,q=0,m).
\end{equation}
Using the same trick as before we see that $\tilde{F}_{\nu}(p)$ can be rewritten as 
\begin{equation}
  \label{rewritting}
  \tilde{F}_{\nu}(p)=-\frac{m\frac{\p }{\p
  m}(G_{2,1}^R(p,-p,q=0,m)/G_2^R(p,m)}{(G_{2,1}^R(p,-p,q=0,m)/G_2^R(p,m))}={Z_2}^{-1}m\frac{\p Z_2}{\p m},
\end{equation}
which shows that $\lim_{p \rightarrow \infty}\tilde{F}_{\nu}(p)=0$, again because $Z_2$ is
a constant in this limit. The equation \eqref{F2gamma2} shows that
$\tilde{F}_{\nu}(s)$ is analytic in $S(\pi-\epsilon')$ for all
$\epsilon'\in (0,\pi)$. This follows from writing
$G_{2,1}^R=G_{2,1}Z_1^{-1}Z_2^{-1}$ and using that 
$G_{2,1}(p,-p,q=0)=-\frac{\p}{\p m_0^2}G_2(p)$ and then 
rewriting in terms of renormalised quantities and doing the spectral 
representation. $\tilde{F}_{\nu}(s)$ then satisfies 
\begin{equation}
  \label{limitsF2}
\tilde{F}_{\nu}(s)\rightarrow \left\{ 
\begin{array}{ll}
2-\tfrac{1}{\nu} & \mbox{for }s\rightarrow 0_{+}, \\
0 & \mbox{for }s\rightarrow \infty. 
\end{array} \right.  
\end{equation}
The functions $\tilde{F}_{\eta}(g)$ and $\tilde{F}_{\nu}(g)$ have been calculated up to an amazing 7 
loops in the coupling constant $g$.  We will follow the
convention of \cite{murray} where $\tilde{g}=\tfrac{3}{16\pi}g$ 
and $\tilde{\beta}(\tilde{g})=\frac{3}{16\pi}\beta(g)$ giving $\pm 1$ as the
first two coefficients in the $\tilde{\beta}$ function; we will use $\beta,
g$ to denote $\tilde{\beta},\tilde{g}$ below to simplify the notation.
The results are that \cite{murray} 
\begin{equation}
  \label{gamma1sok}
\begin{split}
  \tilde{F}_{\eta}(g)\ =\ & 0.0109739369 g^2+0.0009142223 g^3+0.0017962229g^4\\ &\ 
-0.0006536980g^5 +0.0013878101 g^6-0.001697694g^7,\\
  \tilde{F}_{\nu}(g)\ =\ &\tfrac{1}{3} g-0.0631001372 g^2+0.0452244754g^3-0.0377233459 g^4 \\ &+
0.0437466494 g^5 -0.0589756313 g^6+0.09155179g^7,\\ 
 \beta (g)\ =\ & -g+g^2-  0.4224965707g^3+0.3510695978g^4  \\ &
  -0.3765268283g^5   +0.49554751g^6-0.74968893g^7,
\end{split}
\end{equation}
where the beta-function has been calculated up to 6-loops \cite{murray}.
The perturbative series for $\varphi^4$ theory was in \cite{osterwalder} 
shown to be Borel summable in the coupling 
of order $k=1$, so we set $F(s)=\tilde{F}(s^1)$. 
In \cite{BGZJ} it was
shown using the idea from \cite{lipatov} that the coefficients of the
perturbative Green functions $f(g)=\sum_n f_n g^n$ for large $n$ are given as
\begin{equation}
  \label{largek}
  f_n=c(-\tilde{b})^n\Gamma(1+b_l+n)(1+O(1/n))
\end{equation}
where $\tilde{b},b_l,c$ were calculated in
\cite{BGZJ,BP}. $\tilde{b}=0.14777422$ and it follows that the
convergence radius of the Borel transform of $f(g)$ is given by $b_s=1/\tilde{b}$
with a singularity at $-1/\tilde{b}$. 
When $b_l>0$ we want to perform a Borel-Leroy transform of order $k$, 
which replaces a formal series $\hat{f}=\sum_n f_n z^n$ with $\sum_n (f_n/\Gamma
(1+b_l+n/k)z^n$. Instead of \eqref{newcontour} we then define 
\begin{equation}
  \label{leroydef}
  I(\rho)=\frac{\Gamma(1+b_l)}{\rho^{b_l}}\ \frac{1}{2\pi
  i}\int_{\tilde{C}} \frac{ds}{s}s^{b_l} e^{\rho/s}F(s),
\end{equation}
this function will then have the limits $\lim_{\rho \rightarrow \infty}I(\rho)=F_{IR}$ and $\lim_{\rho \rightarrow 0}I(\rho)= F_{UV}$ as before.

We introduce the scale parameter $s$ via the exact running coupling
$\bar{g}(s)$ ($s\rightarrow 0$ in the ultra-violet) then we know that 
$F_{\eta}(\bar{g}(s))$
and $F_{\nu}(\bar{g}(s))$ are asymptotic series in
$\bar{g}(s)$. We now approximate the exact running coupling $\bar{g}(s)$ 
with the solution to 
\begin{equation}
  \label{running1}
  s\frac{d\bar{g}_n(s)}{ds}=-\beta_n(\bar{g}_n)
\end{equation}
with the boundary condition $\bar{g}'_n(0)=1$, 
where $\beta_n(g)$ is the perturbative $\beta$ function to the nth order. 
$F_{\eta}(\bar{g}_n(s))$, $F_{\nu}(\bar{g}_n(s))$ 
are then asymptotic series in $s$,
which we will write as $F_j(s)=\sum_l (F_j)_l s^l$ for $j=\eta, \nu$.
The Borel-Leroy transform given by
\eqref{leroydef} becomes
\begin{equation}
  \label{f1borelsum}
  I(\rho,b_l,k)=\sum_{l=1}^{n+1} \frac{(F_j)_l\ \rho^l}{\Gamma(1+b_l+l/k)}.
\end{equation}
As we are only working with a truncated series we will set $k=1$ (only
infinitesimally different from $k=1+\delta,\ \delta\ll 1$), note we are
summing up to $n+1$ as the nth order $\beta$ function has $n+1$ terms, and
$F_j$ are known to the $(n+1)$th order. 

To find the critical exponents $\nu, \eta$ we will here do an analytical
continuation in the Borel transform (which we do by a conformal mapping)
followed by a Pad\'{e} approximation. The details of this calculation are given
in the appendix. 
We get the following estimates\footnote{
The error in $\nu$ is seen to be larger than the one in $\eta$, in contrast to
the errors form other methods. The reason for this is that we have chosen a
conservative estimate where we have averaged over the results with to
different $\beta$-functions (using its alternating behaviour) as explained in
the appendix. The two $\beta$-functions differ only in the 7th order term, and
here the $F_\eta$ term is about a factor 100 smaller than the $F_\nu$ term
resulting in less sensitivity to this averaging.
The errors obtained could be 
lowered e.g.\ using one of the techniques from
\cite{LGZJ} or as in \cite{BNM} using the information about the pole at
$r_c$ to obtain an extra order in the Pad\'{e} table, but the aim here has
not been to get a very low error, but to apply the method described above to
a well known example.}
\begin{equation}
  \label{nuest}
  \nu=0.625\pm 0.004
\end{equation}
and 
\begin{equation}
  \label{etaest}
  \eta=0.0315\pm 0.0020.
\end{equation}
These numbers should be compared with  $\nu =0.6304\pm 0.0013$ \cite{RGZJ}, 
$\nu =0.6300\pm 0.0015$ \cite{LGZJ}, $\nu =0.6290\pm 0.0025$ ($\epsilon$
expansion \cite{RGZJ}), $\nu=0.6289\pm 0.0008$ (Monte Carlo)  
and for the other exponent
$\eta=0.0355\pm 0.0025$ \cite{RGZJ}, $\eta=0.032\pm 0.003$ \cite{LGZJ},
$\eta=0.0360\pm 0.0050$ ($\epsilon$ expansion \cite{RGZJ}), 
$\eta=0.0374\pm 0.0014$ (Monte Carlo), $\eta=0.0347\pm 0.001$ (strong
coupling \cite{kleinert}), most of these numbers are taken from  
\cite{RGZJ}.

In the usual evaluation one determines the value of the critical coupling
$g_c$ and then evaluates the re-summed series at this point. The values then
becomes very dependent on the estimate of $g_c$. One advantage of 
our method is that we do not have to estimate $g_c$, likewise we do not have
to perform the Laplace integral, but instead the critical point is reached 
taking the limit in the Borel transform. 
The disadvantage of this method is that we do not have specific knowledge of the 
quantities $r_c, b_l$
governing the asymptotic behaviour for the transformed series in the
scale parameter. In the usual case of the perturbative expansion in
the coupling these parameters are obtained from estimates of the higher order 
behaviour of perturbation theory.
 
\section{Conclusion}
We have described a general method of obtaining the infra-red limit of a physical
quantity as an integral in the ultra-violet region. This was done using the analytic
structure of Green functions in a complex scale parameter and by moving all
scale dependence into the running coupling. The infra-red limit is then given
as the limiting value of the Borel transform in the scale 
of the physical quantity. Changing the way in which the scale is introduced
amounts to changing the order in the Borel transform.

We have tested this on two examples where the infra-red limit is well known, namely
for the central charge of the perturbed unitary minimal models and the
critical exponents of $\varphi^4$ theory in three dimensions. 
For the perturbed minimal models we showed how an approximation method can be
obtained for calculating the central charge, and we get approximations close
to the exact values already at one loop by using the largest domain of analyticity
implied by spectral decomposition. For the
$\varphi^4$ theory our estimates of the critical exponents are within the 
errors of other, more elaborate approximations.\\
\ \\
\ \\
{\large \bf Acknowledgement}\\
Lars Kj\ae rgaard acknowledges a research grant from the Danish Research
Academy. We would like to thank R.\ Guida for giving us the reference to
Antonenko and Sokolov in \cite{murray}.

\appendix
\section{Pad\'{e} approximation}
The Borel
transform is analytic in a sector $S(\epsilon'')$ where
$\epsilon''\ll 1$ \cite{balser}. 
The series given by \eqref{f1borelsum} will have a pole at
the negative real axis for some value $\rho=-r_c$ ($r_c>0$) 
determining its convergence radius.
We will now analytically continue $I(\rho)$ given by \eqref{f1borelsum} 
doing the conformal transformation (as in \cite{parisibook})
\begin{equation}
  \label{conftransf}
  t=\frac{\rho}{r_c+\rho}, \ \rho\notin (-\infty,-r_c], \ \ \ \ 
  \rho=\frac{r_c t}{1-t},
\end{equation}
so that $\tilde{I}(t,b_l)=I(\tfrac{r_c t}{1-t},b_l)$. If we write
\eqref{f1borelsum} as $I(\rho)=\sum_l I_l \rho^l$ then $\tilde{I}(t)$ is
given by the series
\begin{equation}
  \label{confborelsum}
\tilde{I}(t)=\sum_{l=1}^{n+1}\tilde{I}_l\ t^l,\ \
\tilde{I}_l=\sum_{m=1}^{l}I_m r_c^m\binom{l-1}{m-1}.
\end{equation}
If all poles of $I(\rho)$ lies in $(-\infty,-r_c]$, as argued in
\cite{LGZJ,parisibook}, then $\tilde{I}(t)$ given by 
this series is convergent for $t\in [0,1)$. 
We want to evaluate this expression at the convergence radius $t=1$ (where
$\tilde{I}(t)$ is regular), to do this we will use Pad\'{e} approximants
\cite{pade,BNM}.
In Pad\'{e} approximation a function is approximated by a rational
function, called $[L/M]$, with polynomials of degree $L$ and $M$ 
in the numerator and denominator (and the constant in the denominator is $1$). 
If the $n$ first terms of the function is known then these
coefficients are matched with the coefficients of the polynomials where
$0<L+N\le n$. One then forms the Pad\'{e} table with entries $[L/M]$. The
Pad\'{e} approximation is based on the conjecture that there is a subsequence
of diagonal Pad\'{e} approximants $[L/L]$ which converge uniformly to the function,
and this conjecture has shown to hold in practice. 
The diagonal Pad\'{e} approximants are conformally invariant and are
therefore independent of $r_c$ above. Generally we should use Pad\'{e}
approximants close to the diagonal. Note that according to the Pad\'{e}
conjecture one should still expect convergence of the Pad\'{e} approximants
even if there are a finite number of poles within $|t|<1$, 
which is the case if not all poles of $I(\rho)$ are in
$(-\infty,-r_c]$.

We also have to determine the 
values of $r_c$ and $ b_l$. Let us first note that
from the boundary condition $\bar{g}_n'(0)=1$ we have that
$\bar{g}_n(s)=s+O(s^2)$ and in $F_j(s)=\sum_l (F_j)_l s^l$ we thus have that 
$(F_j)_l=(\tilde{F}_j)_l+\cdots$ where $(\tilde{F}_j)_l$ 
is the $l$th coefficient of
$g$ in $F_j(g)$,  this means that 
the convergence radius of the transformed series cannot be larger than the
convergence radius of the series in the coupling $g$, i.e.\ 
$0<r_c\le b_s=1/\tilde{b}$.  
Also we would suspect
that $b_l \sim {b'}_l$, where $b'_l$ is the Leroy parameter of the series in
$g$. 
If $b_l$ is chosen too large we will divide by more than the actual
asymptotic increase in \eqref{confborelsum} 
and the approximation value will be to small, if $b_l$
is chosen too small we expect to get a poor convergence in the
Pad\'{e} table. 
In the same way we expect the approximate value 
to be too small if $r_c$ is chosen smaller than the actual convergence radius 
because 
the values we are calculating are
increasing from zero. This is the behaviour we see in the tables and we
estimate $r_c$ and $b_l$ to be respectively the largest and smallest value so
that there is convergence in the Pad\'{e} table. 

From the asymptotic form of the $\beta$ function \eqref{largek} 
it follows that it is
alternating, and an approximation to a truncated alternating asymptotic 
series $\sum_{l=1}^n f_l$ is to use the series $\sum_{l=1}^n\tilde{f}_l$
where $\tilde{f}_l=f_l$ for $l< n$ and $\tilde{f}_n=f_n/2$ \cite{dingle}. We
have also obtained the approximation where the perturbative $\beta$-function 
to the nth order (for $n=5$ and $n=6$) is approximated in this way.

Using the criteria mentioned above, for the choice of $r_c, b_l$  
we get the following Pad\'{e} table\footnote{The numbers marked with a * has
  a pole close to or in the interval $(0,1)$.} 
\begin{equation}
\begin{pmatrix} \cdot & \cdot& 0.407663 & 0.405138 & 0.386709 & 
   0.396082 &  \\ \cdot & 0.362383^* & 0.405248 & 0.408042^* & 
   0.392824 &  &  \\ 0.414164 & 0.395111 & 0.392686 & 
   0.395422 &  &  &  \\ 0.377688 & 0.392075 & 
   0.394118 &  &  & &  \\ 0.4843^* & 
   0.403122 &  &  &  &  &  \\ 
   0.406194 & & &  &  &  & 
   \end{pmatrix}
\end{equation}
for the function $F_{\nu}(s)$ leading to  $F_{\nu,IR} = 0.402  \pm 0.007$,
where the error is the inter-tabular error for the points chosen (which is
here larger than the Baker-Hunter error \cite{BH}), and $r_c=b_s,\ b_l=2.4$. 
Using the $\beta$ function with $1/2$ times the last coefficient we get 
the Pad\'{e} table (we get the best convergence for the 5.\ order expression)
\begin{equation}
  \label{padet}
\begin{pmatrix}\ \ \ \cdot \ & \cdot & 0.407663 & 
   0.405138 & 0.404239 &  \\ \ \ \ \cdot \ & 
   0.362383^* & 0.405248 & 0.403736 &  &  \\ 
  \ \ \ \cdot \ & 0.395111 & 0.404338 &  &  &  \\ 
 \ \ \ \cdot \  & 0.400335 &  &  &  &  \\ 
 \ \ \ \cdot \  &  &  &  &  &  
\end{pmatrix},
\end{equation}
giving $F_{\nu,IR}=0.404\pm 0.004$,
again with $r_c=b_s$ and $b_l=2.4$.
Averaging over these two we get for the critical exponent
$\nu=0.626 \pm 0.003$.
At this value of $b_l$ and $r_c$ we have the best convergence in the Pad\'{e}
table, a more conservative estimate of $F_{\nu,IR}$ is obtained by varying $r_c$
and $b_s$ in a region around these values and then sample the highest and
lowest value for which there is some convergence in the Pad\'{e} 
table\footnote{This variation also have to be performed when expanding
in the coupling as the asymptotic behaviour might not have been reached for
the low number of terms available. 
In \cite{Mudrov} it was argued that the results evaluated using 
Borel-Leroy transforms were very stable under
variation of $b_l$ and $b_s$ in a wide range around the exact asymptotic
values, and we see similar behaviour here.}.
This gives $F_{\nu,IR}=0.40\pm 0.01$ leading to 
\begin{equation}
  \label{nuigen}
  \nu=0.625\pm 0.004.
\end{equation}
For the other exponent $\eta$ we get approximately the same Pad\'{e} tables
using either the $\beta$-function as given in 
\eqref{gamma1sok} 
or with a half times the last
term, the convergent Pad\'{e} table becomes 
\begin{equation}
  \label{padetableeta}
\begin{pmatrix} \cdot & \cdot &\cdot  &\cdot  &\cdot  &\ \ \cdot\ \    \\ \cdot
  &\cdot  &  
   0.0291884 & 0.0321865 & 0.0326601 &    \\ 
  \cdot  & 0.0271003 & 0.0489937^* & 0.0327393 &  & 
      \\ 0.0323937 & 0.0310551 & 0.031359 & 
    &  &    \\ 0.0321239 & 
   0.0312882 & &  & &   
    \\ 0.0312289 &  &  & & 
    &    
\end{pmatrix},
\end{equation}
and we obtain the critical exponent $\eta=0.0319 \pm 0.0010$,
where $r_c=b_s$ and $b_l=1.8$. Again varying around these values gives the
estimate
\begin{equation}
  \label{etany}
  \eta=0.0315\pm 0.0020.
\end{equation}


\begin{thebibliography}{99}

\bibitem{lkpm}
L.~Kj\ae rgaard and P.~Mansfield, \emph{Calculating the infrared
  central charges for perturbed minimal models: improving the rg
  perturbation}, {\it Phys.\ Lett.\ } {\bf B 462} (1999) 103  [hep-th/9905140].

\bibitem{cappelli}
A.~Cappelli, D.~Friedan and J.I. Latorre, \emph{C theorem and spectral
  representation},  {\it Nucl.\ Phys.\ } {\bf B 352} (1991) 616.

\bibitem{cappellipert}
A.~Cappelli and J.I. Latorre, \emph{Perturbation theory of higher spin
  conserved currents off criticality}, {\it Nucl.\ Phys.\ } {\bf B 340} 
(1990) 659;\\
and in {\it Recent developments in Conformal Field
  Theories}, Trieste Conference, World Scientific 1990.

\bibitem{zamolodchikov}
 A.B.\ Zamolodchikov, \emph{`Irreversibility' of the flux
   of the renormalization group in a 2D field theory}, {\it Sov.\ Phys.\  
JETP Lett.\ } {\bf 43} (1986) 730.

\bibitem{francesco}
 P.D.\ Francesco, P.\ Mathieu and D.\ S\'{e}n\'{e}chal,
   \emph{Conformal field theory}, Springer 1996.

\bibitem{stegun}
 {\it Handbook of Mathematical Functions},
M.\ Abramowitz and I.A.\ Stegun eds., Ninth Printing, Dover 1970. 

\bibitem{laessig}
 M.\ L\"{a}ssig, {\em Geometry of the renormalization group
   with an application in two dimensions}, {\it Nucl.\ Phys.\ } {\bf B 334} (1990) 652.


\bibitem{zam2}
A.B. Zamolodchikov, \emph{Renormalization group and perturbation theory near
  fixed points in two-dimensional field theory}, {\it Sov.\ J.\ Nucl.\ Phys.\ } {\bf 46} 
 (1987) 1090.

\bibitem{cardy2}
A.W.W. Ludwig and J.L. Cardy, \emph{Perturbative evaluation of the conformal
  anomaly at new critical points with applications to random systems}, 
{\it Nucl.\ Phys.\ } {\bf B 285} (1987) 687.

\bibitem{intezam} 
A.~Zamolodchikov, \emph{Integrable field theory form conformal field
theory}, \emph{Adv.\ Stud.\ Pure Math.\ } {\bf 19} (1989) 641.

\bibitem{watts} P.~Mathieu and G.~Watts,
\emph{Probing integrable perturbations of conformal theories using singular
vectors}, {\it Nucl.\ Phys.\ } {\bf B 475} (1996) 361 [hep-th/9603088].


\bibitem{cardysum} 
J.L.~Cardy, \emph{The central charge and universal combinations of amplitudes
in two-dimensional theories away from criticality}, 
{\it Phys.\ Rev.\ Lett.\ } {\bf  60}  (1988) 2709.

\bibitem{dotsenko}
 Vl.S.\ Dotsenko and  V.A.\ Fateev, \emph{Operator algebra of
   two dimensional conformal field theory with central charges $c\leq 1$},
{\it Phys.\ Lett.\ } {\bf B 154} (1985) 291;\\
Vl.S.\ Dotsenko, Lectures on conformal field theory, in
  {\em  Conformal field theories and solvable lattice models},  M.\
   Jimbo, T.\ Miwa, A.\ Tsuchiya eds., Academic Press 1988.

\bibitem{alzam}
A.B. Zamolodchikov, \emph{From tricritical ising to critical ising by
  thermodynamic bethe ansatz},  {\it Nucl.\ Phys.\ } {\bf B 358} (1991) 524.

\bibitem{morris}
T.R. Morris, \emph{Properties of derivative expansion approximations to the
  renormalization group},  
{\it Int.\ J.\ Mod.\ Phys.\ } {\bf  B 12} (1998) 1343 [hep-th/9610012];\\
R.~Neves, Y.~Kubyshin and R.~Potting, \emph{Polchinski erg equation and 2d
  scalar field theory},  [hep-th/9811151].

\bibitem{balser}
 W.\ Balser, {\em From Divergent Series to Analytic
    Functions}, LNM 1582, Springer 1994.

\bibitem{parisibook}
 G.\ Parisi, {\em Statistical Field Theory}, Addison
  Wesley 1988.

\bibitem{parisipaper}
 G.\ Parisi, \emph{Field theoretical approach to second
    order phase transitions in two and three dimensional systems}, \emph{J.\
  Stat.\  Phys.\ }{\bf 23} (1980) 49. 

\bibitem{IogD}
 C.\ Itzykson and J.M.\ Drouffe, {\em Statistical field theory},
  vol.\ 1, Cambridge University Press 1989.

\bibitem{RGZJ}
 R.\ Guida and J.\ Zinn-Justin, \emph{Critical Exponents of the
N-vector model}, {\it J.\ Phys.\ }  {\bf A 31}  (1998) 8103, [cond-mat/9803240].

\bibitem{murray}
 D.B.\ Murray and B.E.\ Nickel, unpublished, results reported
  in~\cite{RGZJ}. The 6 loop calculations were first reported in 
G.A. Baker, B.G. Nickel, M.S. Green and D.I. Meiron, \emph{Ising model
  critical indices in three-dimensions from the callan-symanzik equation},
{\it Phys.\ Rev.\ Lett.\ } {\bf 36}  (1976) 1351;\\
It has been calculated to 6 loops for arbitrary $N$ in
S.A. Antonenko and A.I. Sokolov, \emph{Critical exponents for 3d
  $O(n)$-symmetric model with $n > 3$},  
{\it Phys.\ Rev.\ } {\bf E 51}  (1995) 1894 [hep-th/9803264].

\bibitem{osterwalder}
 J.P.\ Echmann, J.\ Magnen and R.\ S\'{e}n\'{e}or, \emph{Decay properties
and Borel summabilities for the Schwinger functions in $P(\Phi)_2$
theories}, 
{\it Comm.\ Math.\ Phys.\ } {\bf  39}  (1975) 251;\\ 
J.\ S.\ Feldman and K.\ Osterwalder,  \emph{The Wightman axioms and the mass gap for weakly coupled
($\phi^4$) in three-dimensions quantum field theories}, 
{\it Ann.\ Phys.\  (NY)} {\bf  97} (1976) 80;\\
J.~Magnen and R.~S\'{e}n\'{e}or, \emph{Phase space cell expansion and borel summability
  for the euclidean $\phi^4$ in three-dimensions theory},  
{\it Comm.\ Math.\ Phys.\ } {\bf 56}  (1977) 237.

\bibitem{BGZJ}
E.~Br\'ezin, J.C.~L. Guillou and J.~Zinn-Justin, \emph{Perturbation theory at
  large order, 1. The $\phi^{2n}$ interaction},  
{\it Phys.\ Rev.\ } {\bf  D 15}  (1977) 1544, 1558.

\bibitem{lipatov}
 L.N.\ Lipatov,  \emph{Calculation of the Gell-Mann-Low function in scalar 
theories with strong nonlinearity}, 
{\it Sov.\ Phys.\ JETP lett.\ } {\bf  24} (1976) 157.

\bibitem{BP}
 E.\ Br\'{e}zin and G.\ Parisi, \emph{J.\ Stat.\ Phys.\ }{\bf 19} (1978) 269.

\bibitem{LGZJ}
J.C.~L. Guillou and J.~Zinn-Justin, \emph{Critical exponents from field
  theory},  
{\it Phys.\ Rev.\ } {\bf B 21}  (1980) 3976.

\bibitem{kleinert}
H.~Kleinert, \emph{Seven-loop critical exponents from strong-coupling
$\phi^4$-theory in three dimensions}, 
{\it Phys.\ Rev.\ } {\bf D 60}  (99) 085001 [hep-th/9812197].

\bibitem{pade}
 G.\ A.\ Baker, Jr.\ and P.\ Graves-Morris, {\em Pad\'{e}
    approximants}, 2.\ ed., Cambridge University Press 1996. 

\bibitem{BNM}
J.G.A.~Baker, B.G. Nickel and D.I. Meiron, \emph{Critical indices from
  perturbation analysis of the Callan-Symanzik equation},  
{\it Phys.\ Rev.\ } {\bf B 17} (1978) 1365.

\bibitem{dingle}
 R.B.\ Dingle, {\em Asymptotic expansions, their derivation
    and interpretation}, Academic Press 1973.

\bibitem{Mudrov}
A.I. Mudrov and K.B. Varnashev, \emph{New approach to summation of
  field-theoretical series in models with strong coupling},
  [hep-th/9811125];\\
\emph{Three-loop renormalization group analysis of a complex model with
stable fixed point: Critical exponents up to $\epsilon^3$ and
$\epsilon^4$}, 
{\it Phys.\ Rev.\ } {\bf B 57}  (1998) 3562 [cond-mat/9712007]. 


\bibitem{BH}
 G.A.\ Baker Jr., and D.L.\ Hunter, \emph{Methods of
series analysis, 1. Comparison of current methods used in the theory 
of critical phenomena}, 
{\it Phys.\ Rev.\ } {\bf B 7} (1973) 3346,\\
 \emph{Methods of series analysis, 2. Generalised and extended methods
with application to the Ising model}, 
{\it Phys.\ Rev.\ } {\bf B 7} (1973) 3377.

\end{thebibliography}
\end{document}